\documentclass[a4paper,12pt]{article}

\usepackage{jheppub}

\title{\boldmath What's the Matter with 3D Gravity?}

\abstract{We revisit the problem of minimally coupling matter to Einstein gravity in three dimensions with negative cosmological constant. By working in the worldline formalism, we construct a classical phase space on an initial time surface $\Sigma$, which we quantize using geometric quantization. States in the Hilbert space correspond to Virasoro conformal blocks with operators of conformal weight $h<c/24$. As an application of our formalism, we compute the partition function on thermal $\text{AdS}_3$ through equivariant localization. Our answer reproduces the AdS$_3$ Wilson spool and agrees with the known one-loop result. It further serves as a conjecture for the value of the path integral of gravity minimally coupled to a massive scalar field in thermal $\text{AdS}_3$ to all orders in $G_N$.}

\author[1]{Robert Bourne,}
\author[2]{Jackson R. Fliss,}
\author[1]{Bob Knighton}
\affiliation[1]{Department of Applied Mathematics \& Theoretical Physics, University of Cambridge,\\
Wilberforce Road, Cambridge CB3 0WA, United Kingdom}
\affiliation[2]{Physique Th\'eorique et Math\'ematique, Universit\'e Libre de Bruxelles \&\\
International Solvay Institutes, CP 231, 1050 Bruxelles, BE}
\emailAdd{rjb260@cam.ac.uk}
\emailAdd{jackson.fliss@ulb.be}
\emailAdd{rik23@cam.ac.uk}

\usepackage{bm} 
\usepackage[dvipsnames]{xcolor}
\definecolor{green_maf}{RGB}{28, 166, 46}
\definecolor{blue_mrg}{RGB}{12, 143, 145}
\definecolor{detail}{RGB}{110,110,110}
\usepackage{amsmath} 
\usepackage{nccmath} 
\usepackage{amssymb} 
\usepackage{amsthm} 
\usepackage{mathtools} 
\usepackage[utf8]{inputenc} 
\usepackage{braket}
\usepackage{enumerate}
\usepackage[many]{tcolorbox}
\usepackage[inline]{enumitem}
\usepackage{comment}
\usepackage{soul} 
\usepackage{tikz}
\usepackage{csquotes}
\usepackage{mathrsfs}
\usepackage{tensor}
\usepackage{dsfont}
\usepackage{slashed}

\newtcolorbox{empheqboxed}{colback=gray!30, 
 colframe=white,
 width=\textwidth,
 sharpish corners,
 top=-2mm, 
 bottom=0pt
}

\hypersetup{
		pdfencoding=unicode,
		colorlinks=true,
		urlcolor=Maroon,
		linkcolor=RoyalBlue,
		citecolor=Maroon,
		pdfstartview=FitH,
		linktocpage=true
}

\usetikzlibrary{calc,decorations.markings}
\usetikzlibrary{decorations.pathmorphing}
\usetikzlibrary{shapes,backgrounds,shapes.misc}
\usetikzlibrary{fadings}
\usetikzlibrary{cd}
\usetikzlibrary{decorations.pathreplacing,calligraphy}
\usetikzlibrary{hobby}

\tikzset{
	partial ellipse/.style args={#1:#2:#3}{
		insert path={+ (#1:#3) arc (#1:#2:#3)}
	}
}

\tikzset{
  every overlay node/.style={
    draw=black,fill=white,rounded corners,anchor=north west,
  },
}

\tikzfading
[
  name=fade out,
  inner color=transparent!0,
  outer color=transparent!100
]

\newif\ifdetails
\detailstrue

\newcommand{\dd}{{\rm d}} 

\newcommand\mc[1]{\mathcal{#1}}
\newcommand{\bTr}[2]{\text{Tr}\left(#1\middle|#2\right)}

\newcommand{\Hg}{\mathcal{H}_{\text{grav}}}
\newcommand{\Hgm}{\mathcal{H}_{\text{grav+matter}}}
\newcommand{\Hgn}{\mathcal{H}_{\text{grav,n}}}
\newcommand{\Mg}{\mathcal{M}_{\text{grav}}}
\newcommand{\Mgn}{\mathcal{M}_{\text{grav},n}}
\newcommand{\grav}{\text{grav}}

\newcommand{\gpm}{\text{grav+matter}}

\def\be{\begin{equation}}
\def\ee{\end{equation}}
\def\beq{\begin{equation}}
\def\eeq{\end{equation}}

\begin{document}

\maketitle

\pagebreak
\section{Introduction}

What's the matter with 3D gravity? On the surface, nothing. It is by now well known that pure gravity in low enough dimensions, i.e. $d\leq3$, is both quantizable and renormalizable \cite{Witten:1988hc,Witten:2007kt}. This is owing to the absence of a local propagating on-shell graviton and the related recasting of the theory as a topological quantum field theory (TQFT) \cite{Achucarro:1986uwr}. This fact, paired with the AdS/CFT correspondence, has led to a wealth of insights of the quantum nature of pure gravity and black holes in the presence of negative cosmological constant, $\Lambda$ \cite{Brown:1986nw,Coussaert:1995zp,Barnich:2017jgw, Cotler:2018zff}. Substantial progress was recently made on this front in the development of the {\it Virasoro TQFT} \cite{Collier:2023fwi,Collier:2024mgv}, a precise and consistent treatment of pure 3D gravity with $\Lambda<0$\footnote{The state of the art for $\Lambda>0$ is much more rudimentary, however see \cite{Castro:2011xb,Castro:2012gc,Cotler:2019nbi,Hikida:2021ese,Shyam:2021ciy,Coleman:2021nor,Narovlansky:2023lfz,Batra:2024kjl,Collier:2025lux} for progress utilizing similar ethe.} that allows one to leverage standard techniques of TQFT to compute the quantum gravitational partition function on any hyperbolic three-manifold\footnote{Due to the Mostow rigidity theorem \cite{Mostow1968-im}, in three dimensions this is a topological designation as opposed to a geometric one.} to all orders in Newton's constant, $G_N$. 

So, what's the matter with 3D gravity? Despite the above progress, there are many puzzles about pure gravity left unresolved by its reformulation as a TQFT. Perhaps the most famous of these puzzles is the one posed by Maloney and Witten \cite{Maloney:2007ud} who realized that the canonical partition function of pure AdS$_3$ gravity yields both a potentially negative and continuous density of states for a putative dual CFT. The first issue has been addressed by including certain off-shell configurations \cite{Benjamin:2020mfz,Maxfield:2020ale} while the second is a strong indication that pure 3D gravity is not dual to a single CFT but instead encodes the statistics of an ensemble of theories at large central charge \cite{Belin:2020hea,Cotler:2020ugk,Chandra:2022bqq,DiUbaldo:2023qli,Haehl:2023mhf,Belin:2023efa,deBoer:2023vsm,Jafferis:2025vyp,Boruch:2025ilr}; this interpretation is further bolstered by the existence of both on-shell and off-shell connected topologies, i.e. Euclidean wormholes \cite{Maldacena:2004rf,Yin:2007gv,Cotler:2020ugk}. The upshot of these discussions is that pure 3D gravity is likely a coarse-grained description of a microscopic theory as opposed to one itself. Said another way, while 3D gravity may be rendered UV {\it finite}, it is not UV {\it complete}.

If string theory is a good `top-down' indication towards a UV complete theory of quantum gravity, we know that gravity never comes alone; it comes part and parcel with a tower of massive modes which around a given background manifest themselves as propagating quantum field theories. Indeed the inclusion of matter, at the level of point particles, is the most minimal interpretation of the solutions to the negative density of states proposed in \cite{Benjamin:2020mfz,Maxfield:2020ale,DiUbaldo:2023hkc}. In this paper, we will take a `bottom-up' perspective and ask how one can add matter to 3D gravity, to what degree it's quantizable, and if it can tell us anything about gravity in the UV.

So, what's the {\it matter} with 3D gravity? In spite of lofty string theory motivations, in this paper we take a modest first step and consider a massive scalar field minimally coupled to gravity. Stated as a Euclidean path integral the theory is
\beq\label{eq:Zgpm1}
    Z_{\gpm}=\int\mc Dg_{\mu\nu}\mc D\phi\;e^{-S_\text{EH}[g]}\exp\left(-\frac{1}{2}\int\dd^3x\,\sqrt{g}\left(g^{\mu\nu}\partial_\mu\phi\partial_\nu\phi+m^2\phi^2\right)\right)~,
\eeq
with $S_\text{EH}$ the Einstein-Hilbert action
\begin{equation}
S_{\text{EH}}[g]=\frac{1}{16\pi G_N}\int\mathrm{d}^3x\sqrt{g}\left(R-2\Lambda\right)\,.
\end{equation}
While indeed modest, we might regard \eqref{eq:Zgpm1} as the starting point for perturbation theory for more robust effective field theories of quantum matter coupled to quantum gravity. Regardless, the results of \cite{Benjamin:2020mfz,Maxfield:2020ale} indicate that already at the level of point particles, the backreaction of scalar matter drastically alters the UV interpretation of 3D gravity. While \eqref{eq:Zgpm1} certainly won't render the density of states discrete, explicitly including matter might indicate more fine-grained statistical features a CFT ensemble should possess to abide bulk backreaction. As an additional motivation, one might suspect the backreaction of minimally coupled matter could be sufficient for extending the technology of the Virasoro TQFT for computing quantities on nonhyperbolic manifolds. Lastly, from truly bottom-up considerations, \eqref{eq:Zgpm1} is precisely the type of bulk effective field theory one considers in traditional AdS/CFT when discussing single trace CFT operators and it is of its own interest to develop techniques for evaluating $G_N\sim c^{-1}$ corrections to observables. The above motivations aside, we will see that \eqref{eq:Zgpm1}, while modest, contains rich physics.

The problem at hand is both relatively straightforward and extremely subtle. On the one hand, the scalar path integral in \eqref{eq:Zgpm1} is Gaussian and can be, at least formally, integrated. The issue is that the resulting expression is a highly non-linear and non-local functional of the metric. Expanding it as a power series in $G_N$ is essentially a Feynman-diagrammatic expression of the gravitational effective action and completely ignores the computational power (and conceptual insight) afforded to us by pure 3D gravity as a TQFT. Ultimately, this issue stems from the fact that \eqref{eq:Zgpm1} is not a TQFT: $\phi$ is a local propagating degree of freedom. To overcome this inconvenient fact, we take heart from the {\it Wilson spool} program \cite{Castro:2023dxp,Castro:2023bvo,Bourne:2024ded} which expresses the one-loop determinant of minimally coupled matter as a topological line operator inside the gravitational path integral -- schematically
\beq\label{eq:Zgpm2}
    Z_\gpm=\int \mathcal Dg_{\mu\nu}e^{-S_\text{EH}[g]}\exp\mathbb W[g]\equiv \Big\langle\exp\mathbb W\Big\rangle_\grav~.
\eeq
The Wilson spool, $\mathbb W$, is a collection of line operators wrapping all non-trivial cycles of the background topology arbitrarily many times \cite{Bourne:2025azc} and is most naturally expressed in the TQFT formulation of gravity. $\mathbb W$ maintains the topological nature of low-dimensional gravity, and allows for the efficient calculation of quantum gravitational effects, when those tools exist \cite{Castro:2023dxp}. In the context of AdS$_3$ gravity, \eqref{eq:Zgpm2} has been confirmed to tree-level (in a $G_N$ expansion) for all smooth and cusp-free hyperbolic backgrounds \cite{Bourne:2025azc}. Here we wish to go beyond this and provide a structured procedure for evaluating \eqref{eq:Zgpm2} {\it exactly}.

In order to do so, we take orthogonal inspiration from the Virasoro TQFT. Stated differently, our objective is to find a construction of $\mathbb W$ within the Virasoro TQFT. Doing so requires us to revisit the canonical quantization of 3D gravity in the presence of matter. While several aspects of this construction already appear in \cite{Collier:2023fwi}, here we press this construction to realize the entire scalar field theory directly within the Virasoro TQFT, including perturbative and non-perturbative effects of its backreaction. The upshot of this is a concrete construction of the Hilbert space of gravity coupled to matter from which one can, in principle, evaluate $Z_\gpm$ exactly. We put our formalism to the test, showing that for thermal AdS$_3$ our construction nicely reduces to the Wilson spool of \cite{Castro:2023bvo,Bourne:2025azc}, reproduces the known one-loop partition functions of Giombi, Maloney, and Yin (GMY) \cite{Giombi:2008vd} at tree-level, and can be explicitly evaluated to all orders in $G_N$ perturbation theory. Surprisingly, this perturbation theory also tells us about nonperturbative effects as well.

An outline of the paper is summarized as follows. In Section \ref{sec:pure-gravity} we review the basic construction of the gravitational Hilbert space from the K\"ahler quantization of its phase space. We then modify this construction to include matter in Section \ref{sec:including-matter}; inspired by the worldline approach to scalar field theory, we realize the scalar field as a ``gas of defects'' acting on the gravitational phase space. This provides a natural symplectic form and inner product which we quantize. In Section \ref{sec:partition-functions} we show the utility of our construction in computing exact partition functions, relying on index theorems and equivariant localization which we review in that section. As a concrete application, we perform this calculation for thermal AdS$_3$ where we make an explicit connection to the standard Wilson spool and the GMY results. While the construction in Section \ref{sec:including-matter} is in principle nonperturbative, in order to perform this computation, we ignore, as an approximation, nonperturbative effects. Nevertheless our result is exact to all orders in $G_N$ perturbation theory; we show in Section \ref{sec:black-holes} that this can be resummed and in doing so we can deduce nonperturbative contributions to the partition function. Finally in Section \ref{sec:discussion} we conclude with a discussion of our results, open questions, and speculative thoughts.

\section{Quantization of pure gravity}\label{sec:pure-gravity}

Before considering the more general problem of quantizing gravity coupled to matter, let us briefly summarize what is known about the quantization of pure $\text{AdS}_3$ gravity. The concepts, and in particular the discussion of the phase space, will be instrumental in understanding the more general problem which we will return to in Section \ref{sec:including-matter}.

In the first-order formulation of Einstein gravity in three dimensions, the dynamical degrees of freedom are the dreibein $e^a$ and the spin connection $\omega^{ab}$. These can be combined into $\text{PSL}(2,\mathbb{R})$ gauge fields
\begin{equation}\label{eq:CStograv}
A_L^a=\frac{1}{2}\varepsilon\indices{^a_b_c}\omega^{bc}+\frac{1}{\ell}e^a\,,\quad A_R^a=\frac{1}{2}\varepsilon\indices{^a_b_c}\omega^{bc}-\frac{1}{\ell}e^a\,,
\end{equation}
with $\Lambda=-1/\ell^2$. The frame indices are to be identified as $\mathfrak{psl}(2,\mathbb{R})$ indices, and the Einstein-Hilbert action takes the form
\begin{equation}
\begin{split}
S_{\text{EH}}[A_L,A_R]=&\frac{k}{4\pi}\int\text{Tr}\left(A_L\wedge\mathrm{d}A_L+\frac{2}{3}A_L\wedge A_L\wedge A_L\right)\\
&-\frac{k}{4\pi}\int\text{Tr}\left(A_R\wedge\mathrm{d}A_R+\frac{2}{3}A_R\wedge A_R\wedge A_R\right)\,.
\end{split}
\end{equation}
The level $k$ of the Chern-Simons action is given by
\begin{equation}
k=\frac{\ell}{16G_N}\,,
\end{equation}
and is related to the Brown-Henneaux central charge by $c=24k$. The equations of motion derived from the Chern-Simons action, i.e. the flatness of the connections $A_{L,R}$, are equivalent to the three-dimensional Einstein equations in the metric formulation.

\paragraph{The classical phase space:} Although Chern-Simons theory and 3D gravity are equivalent at the level of the classical action, their phase spaces are genuinely different, since not every Chern-Simons connection gives rise to a smooth classical metric. The gravitational phase space is instead identified with a subset of the phase space of Chern-Simons theory.

Let us be more specific and fix an initial time surface $\Sigma$. The phase space of Chern-Simons theory with group $G$, after gauge-fixing, is given by the moduli space $\mathcal{M}_G$ of flat $G$-bundles on $\Sigma$. Thus, the gravitational phase space sits inside two copies of the moduli space of flat $\text{PSL}(2,\mathbb{R})$ bundles \cite{Witten:1988hf}:
\begin{equation}
\Mg\subset\mathcal{M}_{\text{PSL}(2,\mathbb{R})}\times\overline{\mathcal{M}}_{\text{PSL}(2,\mathbb{R})}\,.
\end{equation}
Here, the bar does not represent complex conjugation, but rather emphasizes that the two $\text{PSL}(2,\mathbb{R})$ bundles are independent. The phase space of flat $\text{PSL}(2,\mathbb{R})$ connections is not a connected space, and contains in particular a distinguished connected component, the Teichm\"uller space $\mathcal{T}_{\Sigma}$ of the surface $\Sigma$. The precise embedding of the gravitational phase space inside of the moduli space of flat connections was found in \cite{Krasnov:2005dm,Scarinci:2011np} to be
\begin{equation}
    \Mg\cong\mathcal{T}_{\Sigma}\times\overline{\mathcal{T}}_{\Sigma}\,.
\end{equation}
The symplectic form on the gravitational phase space can be derived from the Chern-Simons action, and is given by
\begin{equation}
\omega=\frac{k}{4\pi}\omega_{\text{WP}}-\frac{k}{4\pi}\overline{\omega}_{\text{WP}}\,,
\end{equation}
where $\omega_{\text{WP}}$ (resp. $\overline{\omega}_{\text{WP}}$) is the Weil-Petersson symplectic form on $\mathcal{T}_{\Sigma}$ (resp. $\overline{\mathcal{T}}_{\Sigma}$).

\paragraph{Quantization:} Since the gravitational phase space is equivalent to two copies of Teichm\"uller space, the quantization of 3D gravity is equivalent to the quantization of $\mathcal{T}_{\Sigma}$ (up to issues involving the mapping class group which we will discuss below). We will first focus on one copy of Teichm\"uller space.

Quantization proceeds as follows \cite{Verlinde:1989ua}. Making use of the K\"ahler structure of Teichm\"uller space, we identify a holomorphic line bundle $\mathcal{L}$ over $\mathcal{T}_{\Sigma}$ whose first Chern class is in the same cohomology class as the symplectic form:
\begin{equation}
c_1(\mathcal{L})=\frac{1}{4\pi}[\omega_{\text{WP}}]\,.
\end{equation}
The Hilbert space on $\Sigma$ is then identified with the space $\text{H}^0(\mathcal{T}_{\Sigma},\mathscr{L})$ of holomorphic sections of $\mathscr{L} = \mathcal{L}^{k}$. These holomorphic sections have an intuitive interpretation in terms of left-moving conformal blocks of 2D conformal field theories. Specifically, a section $\Psi$ of $\mathscr{L}$ is an object which transforms like a 2D CFT partition function on $\Sigma$ of left-moving central charge $c_L=24k$ in the sense that it satisfies all of the conformal Ward identities, although not necessarily crossing symmetry. That $\Psi$ is a holomorphic section means that it only depends on the moduli of the surface in a holomorphic fashion.

Since the full gravitational phase space is two copies of Teichm\"uller space, the Hilbert space will be the tensor product of a left- and right-moving space:
\begin{equation}
\Hg=\mathcal{H}_L\otimes\mathcal{H}_R\,,
\end{equation}
where $\mathcal{H}_L$ ($\mathcal{H}_R$) is the space of holomorphic (antiholomorphic) sections of $\mathscr{L}$ ($\overline{\mathscr{L}}$). Elements of $\mathcal{H}_{\text{grav}}$ are functions of two independent left- and right-moving moduli which transform as CFT partition functions of central charge $(c,c)$ under left- and right-moving Weyl transformations.

\paragraph{The inner product:} The Hilbert space obtained from quantizing Teichm\"uller space comes equipped with a natural inner product, which can be derived rigorously using the measure on $\mathscr{L}$ induced by the Weil-Petersson form \cite{Verlinde:1989ua} or by analogy with string theory \cite{Collier:2023fwi}. In the latter approach, we consider two sections $\Psi_1,\Psi_2$ of $\mathscr{L}$. Their product $\overline{\Psi}_1\Psi_2$ transforms like a CFT partition function of central charge $(c,c)$. From string theory, we know that the natural objects that can be integrated over Teichm\"uller space are those with central charge $(26,26)$. In \cite{Collier:2023fwi}, it was proposed that the correct procedure is to integrate $\overline{\Psi}_1\Psi_2$ against the partition function of \textit{timelike Liouville theory}, $Z_\text{tL}$, which has central charge $c_\text{tL}=26-c$. The resulting inner product is then
\begin{equation}\label{eq:vtqft-inner-product}
\braket{\overline{\Psi}_1,\Psi_2}=\int_{\mathcal{T}_{\Sigma}}Z_{bc}\,Z_{\text{tL}}\,\overline{\Psi}_1\Psi_2\,,
\end{equation}
where $Z_{bc}$ is the partition function of the usual conformal ghost system familiar from string theory.

\paragraph{Large diffeomorphisms and crossing symmetry:} One important difference between the quantization of gravity and that of Chern-Simons theory is that in gravity one must gauge both small and large diffeomorphisms. The Chern-Simons description already includes the gauging of small diffeomorphisms, but we still have to deal with the large diffeomorphisms by hand. The set of all large diffeomorphisms modulo small diffeomorphisms is known as the \textit{mapping class group}, which we denote as $\text{Map}(\Sigma)$ for a given surface.

If we equip $\Sigma$ with a metric $h$, an element of $\text{Map}(\Sigma)$ acts on $h$ in a way that cannot be compensated by a Weyl transformation. As such, $\text{Map}(\Sigma)$ acts naturally on the Teichm\"uller space, and as such the gravitational phase space. Gauging the mapping class group is then achieved by treating two points in phase space that differ by a large diffeomorphism as physically equivalent, so that the `true' gravitational phase space is
\begin{equation}\label{eq:phase-space-map(sigma)}
(\mathcal{T}_{\Sigma}\times\overline{\mathcal{T}}_{\Sigma})/\text{Map}(\Sigma)\,,
\end{equation}
where $\text{Map}(\Sigma)$ acts diagonally on the two copies of Teichm\"uller space. Since the Weil-Petersson form is invariant under large diffeomorphisms, the phase space \eqref{eq:phase-space-map(sigma)} naturally inherits a symplectic structure from $\mathcal{T}_{\Sigma}\times\overline{\mathcal{T}}_{\Sigma}$.

In the quantum theory, there are effectively two ways to go about gauging the mapping class group. We can either gauge before quantization, so that the quantum Hilbert space is found by quantizing \eqref{eq:phase-space-map(sigma)}, or we can gauge after quantization. In the latter approach, physical wavefunctions are sections of the line bundle discussed above which are invariant under the action of $\text{Map}(\Sigma)$. In either case, since large diffeomorphisms act on conformal blocks by crossing transformations, the gravitational phase space after gauging $\text{Map}(\Sigma)$ is precisely the subspace of conformal blocks $\mathcal{H}_{\text{grav}}$ which are crossing-symmetric.


\section{Including matter}\label{sec:including-matter}

We now turn our attention to the quantization of gravity minimally coupled to a non-gravitational field theory. While there are many field theories we could couple to the gravitational field, we will focus on the prototypical example of a single free scalar field of mass $m$. We will comment on how one might generalize this analysis to include multiple particle species, spinning fields, and local interactions in the discussion in Section \ref{sec:discussion}.

As mentioned in the introduction, the gravitational path integral coupled to a massive scalar takes the form
\begin{equation}
Z_{\text{grav}+\text{matter}}=\int\mathcal{D}g\,\mathcal{D}\phi\,e^{-S_{\text{EH}}-S[\phi,g]}\,,
\end{equation}
where $S[\phi,g]$ is the usual Euclidean scalar action
\begin{equation}
S[\phi,g]=\frac{1}{2}\int\mathrm{d}^3x\sqrt{g}\left(g^{\mu\nu}\partial_{\mu}\phi\partial_{\nu}\phi+m^2\phi^2\right)\,.
\end{equation}
In principle, we could quantize this theory in much the same way as we did for pure gravity, first by writing down the classical phase space, and then by constructing the appropriate Hilbert space via canonical quantization. Indeed, the classical phase space for this theory is known \cite{Witten:2022xxp} to take the following simple form. Let $\Phi_{\Sigma}$ denote the space of configurations of $\phi$ on an initial time surface $\Sigma$. Then the gravitational phase space is given by the cotangent bundle
\begin{equation}\label{eq:gravity+matter-witten}
T^*(\text{Met}(\Sigma)\times\Phi_{\Sigma})/\text{Weyl}(\Sigma)\rtimes\text{Diff}(\Sigma)\,,
\end{equation}
where Weyl transformations are taken to act trivially on $\phi$. 

In the case that no matter is present, i.e. when $\Phi_{\Sigma}$ is empty, the phase space \eqref{eq:gravity+matter-witten} can be written in terms of Teichm\"uller space:
\begin{equation}\label{eq:pure-gravity-cotangent}
T^*\mathcal{T}_{\Sigma}/\text{Map}(\Sigma)\,.
\end{equation}
At first glance, this does not look like the phase space \eqref{eq:phase-space-map(sigma)} established from the Chern-Simons formulation. However, the two are equivalent via the ``Mess map'' \cite{Mess:2007abc}, which establishes the symplectomorphism
\begin{equation}
T^*\mathcal{T}_{\Sigma}\cong\mathcal{T}_{\Sigma}\times\overline{\mathcal{T}}_{\Sigma}\,.
\end{equation}
The Mess map explicitly shows that the phase space \eqref{eq:pure-gravity-cotangent} has a K\"ahler structure (that of $\mathcal{T}_{\Sigma}$), and thus quantization can be done via K\"ahler quantization as in Section \ref{sec:pure-gravity}.

What about when we add matter? In this case, there is no analogue of the Mess map that can express the phase space as the product of complex manifolds, and thus it is far from obvious whether \eqref{eq:gravity+matter-witten} carries a K\"ahler structure. Furthermore, the phase space \eqref{eq:gravity+matter-witten} is infinite-dimensional, and thus its quantization is as difficult as the quantization of a non-topological quantum field theory. In short, the two properties that rendered pure 3D gravity easily quantizable (K\"ahler structure and finite-dimensionality of the phase space) have been lost by the inclusion of our matter field.

\subsection{A scalar field as a gas of defects}

The way forward, as we will see, is to instead work in the worldline formalism, whereby we can trade the path integral over $\phi$ for the path integral over worldlines of a relativistic point particle of mass $m$:
\begin{equation}\label{eq:scalar-to-worldline}
\int\mathcal{D}\phi\,e^{-S[g,\phi]}=\exp\left(\int\mathcal{D}x\,e^{-S_{\text{worldline}}[g,x]}\right)\,.
\end{equation}
Let $M$ be the three-manifold of interest. The worldline path integral is over all connected paths $x:\text{S}^1\to M$ and the action is the proper length of the worldline as measured by the background metric $g$:
\begin{equation}
S_{\text{worldline}}[g,x]=\frac{\mu}{\ell}\int\mathrm{d}\tau\,.
\end{equation}
Here $\mu$ is the worldline coupling in AdS units. Semiclassically, one expects $\mu=m\ell$ to hold for heavy fields ($m\gg 1/\ell$). However, due to both the curvature of the background and the coupling of the worldline to gravity, the relationship between $\mu$ and the scalar mass is subject to quantum corrections (see for example \cite{Maxfield:2017rkn}). For the moment we will therefore keep $\mu$ arbitrary and fix it by comparison to the one-loop analysis of Giombi, Maloney, and Yin \cite{Giombi:2008vd} in Section \ref{sec:comparison} as a sort of renormalization condition. Up to potential $\mathcal{O}(G_N)$ corrections, we will find the correct prescription to be
\begin{equation}\label{eq:self_energy_mass_relation}
\mu=1+\sqrt{m^2\ell^2+1}\,,
\end{equation}
which is the value of a dual conformal weight in the AdS/CFT dictionary. The effect of expanding the exponential in equation \eqref{eq:scalar-to-worldline} is that we integrate over all multiparticle worldline configurations, weighted by appropriate symmetry factors to account for Bose-Einstein statistics.

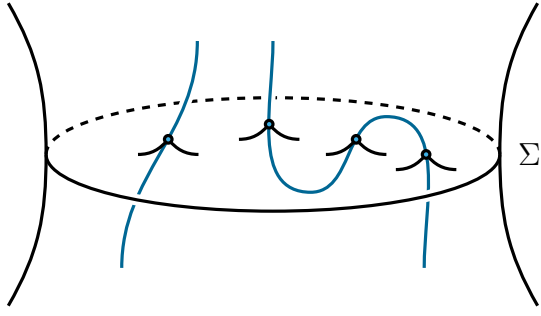
\begin{figure}
\centering
\begin{tikzpicture}
\draw[very thick, dashed] (0,0) [partial ellipse = 0:180:3 and 0.75];
\draw[line width = 0.15cm, white] (-2,-1.5) to[out = 90, in = -90] (-1,1.5);
\draw[very thick, MidnightBlue] (-2,-1.5) to[out = 90, in = -90] (-1,1.5);
\begin{scope}[xshift = 0.115cm]
    \draw[very thick] (-1.5-0.4,0) to[out = 0, in = -130] (-1.5,0+0.2) to[out = -50, in = 180] (-1.5 + 0.4,0);
    \draw[very thick, fill = CornflowerBlue] (-1.5,0+0.2) circle (0.05);
\end{scope}
\draw[line width = 0.15cm, white] (2,-1.5) to[out = 90, in = 0] (1.5,0.5) to[out = 180, in = 0] (0.5,-0.5) to[out = 180, in = -90] (0,1.5);
\draw[very thick, MidnightBlue] (2,-1.5) to[out = 90, in = 0] (1.5,0.5) to[out = 180, in = 0] (0.5,-0.5) to[out = 180, in = -90] (0,1.5);
\begin{scope}[xshift = -0.025cm]
    \draw[very thick] (2.05-0.4,-0.2) to[out = 0, in = -130] (2.05,-0.2+0.2) to[out = -50, in = 180] (2.05 + 0.4,-0.2);
    \draw[very thick, fill = CornflowerBlue] (2.05,-0.2+0.2) circle (0.05);
\end{scope}
\begin{scope}[xshift = 0.1cm]
    \draw[very thick] (1-0.4,0) to[out = 0, in = -130] (1,+0.2) to[out = -50, in = 180] (1 + 0.4,0);
    \draw[very thick, fill = CornflowerBlue] (1,0.2) circle (0.05);
\end{scope}
\begin{scope}
    \draw[very thick] (-0.05-0.4,0.2) to[out = 0, in = -130] (-0.05,0.2+0.2) to[out = -50, in = 180] (-0.05 + 0.4,0.2);
    \draw[very thick, fill = CornflowerBlue] (-0.05,0.2+0.2) circle (0.05);
\end{scope}
\draw[line width = 0.15cm, white] (0,0) [partial ellipse = 0:-180:3 and 0.75];
\draw[very thick] (0,0) [partial ellipse = 0:-180:3 and 0.75];
\draw[very thick] (-3.5,-2) to[out = 60, in = -90] (-3,0) to[out = 90, in = -60] (-3.5,2);
\draw[very thick] (3.5,-2) to[out = 120, in = -90] (3,0) to[out = 90, in = -120] (3.5,2);
\node[right] at (3.1,0) {$\Sigma$};
\end{tikzpicture}
\caption{Point-particle worldlines intersecting a spacelike surface $\Sigma$. The point particles induce curvature singularities (conical defects) on $\Sigma$.}
\label{fig:worldlines}
\end{figure}

The primary benefit of the point-particle picture is that it has an immediate geometric interpretation in terms of the types of metrics we are allowed to integrate over in the gravitational path integral. Consider the contribution of a single worldline which sweeps out the path $\gamma$. Then the equations of motion for the metric can be derived by minimizing the action
\begin{equation}
\frac{1}{16\pi G_N}\int\mathrm{d}^3x\sqrt{g}\left(R+\frac{2}{\ell^2}\right)+\frac{\mu}{\ell}\int_{\gamma}\mathrm{d}\tau\,,
\end{equation}
where $\tau$ is the proper distance along the curve $\gamma$. Varying the action gives the modified (traced) Einstein equations
\begin{equation}\label{eq:EinEQwithdefect}
R+\frac{6}{\ell^2}=-\frac{16\pi G_N\mu}{\ell}\int_{\gamma}\mathrm{d}\tau\,\delta^{(3)}(x(\tau),x)\,.
\end{equation}
Solutions to this equation are locally-$\text{AdS}_3$ manifolds with curvature singularities localized along the particle worldlines, see Figure \ref{fig:worldlines}. In three dimensions specifically, these are conical singularities with defect angle
\begin{equation}\label{eq:angletomass}
\theta=\frac{8\pi G_N\mu}{\ell}\,.
\end{equation}
We can associate to this defect angle a chiral ``conformal weight'' 
\beq\label{eq:conformal-weight-main}
    h\equiv k(1-\alpha_\theta^2) = \frac{\mu}{2}-\frac{G_N\mu^2}{\ell}~,\qquad \alpha_\theta=1-\frac{\theta}{2\pi}~.
\eeq
At this point, $h$ is simply a reparameterization of the mass parameter, however we shall see that in scenarios when $\Sigma$ possesses an asymptotic boundary, $h$ is the chiral conformal weight of a dual scalar primary operator through the standard AdS/CFT dictionary.

Given an initial time surface $\Sigma$ in $M$, as well as a collection of particle worldlines, there will be a well-defined `particle number' given by the number of points on $\Sigma$ that intersect some worldline. For a fixed particle number, $n$, the gravitational phase space is known to be \cite{Bonsante:2007abc}
\begin{equation}
    \Mgn=(\mathcal{T}_{\Sigma,n,\theta}\times\overline{\mathcal{T}}_{\Sigma,n,\theta})/\text{Map}(\Sigma,n)\,.
\end{equation}
The space $\mathcal{T}_{\Sigma,n,\theta}$ is the Teichm\"uller space of \textit{cone surfaces}, and $\text{Map}(\Sigma,n)$ is the mapping class group of $\Sigma$ with $n$ identical punctures.\footnote{More generally we write $\mathcal{T}_{\Sigma,\{\theta_1,\ldots,\theta_n\}}$ for the Teichm\"uller space of cone surfaces with angles $\theta_1,\ldots,\theta_n$.} Both $\mc T_{\Sigma,n,\theta}$ and $\text{Map}(\Sigma,n)$ have intuitive physical interpretations suggested precisely by worldline formalism. The Teichm\"uller space $\mc T_{\Sigma,n,\theta}$ is the space of hyperbolic metrics with $n$ cone singularities with defect angle $\theta$ which arise as classical solutions to \eqref{eq:EinEQwithdefect} restricted to $\Sigma$ (suitably modified to account for $n$ point masses) up to small diffeomorphisms:
\begin{equation}
\mathcal{T}_{\Sigma,n,\theta}=\Set{g_{ij},x_1,\ldots,x_n|R^{(2)}_g(x)=-2-2\theta\sum_{i=1}^{n}\delta^{(2)}(x,x_i)}/\text{Diff}_0(\Sigma,n)\,,
\end{equation}
where $g_{ij}$ and $R^{(2)}_g$ are a metric and scalar curvature on $\Sigma$. The mapping class group is the space of large diffeomorphisms:
\begin{equation}
\text{Map}(\Sigma,n)=\text{Diff}(\Sigma,n)/\text{Diff}_0(\Sigma,n)\,.
\end{equation}
Since we are working with a theory of identical particles, we allow elements of $\text{Map}(\Sigma,n)$ to permute the marked points $x_1,\ldots,x_n$. A fact that will be useful for us later is that $\text{Map}(\Sigma,n)$ is an extension of the usual `pure' mapping class group $\text{PMap}(\Sigma,n)$ (i.e. the group of large diffeomorphisms which do not permute the defects) by the symmetric group $S_n$. More concretely, $\text{Map}(\Sigma,n)$ sits at the center of the exact sequence
\begin{equation}
1\to\text{PMap}(\Sigma,n)\to\text{Map}(\Sigma,n)\to S_n\to 1\,.
\end{equation}

As in the case of pure gravity described in Section \ref{sec:pure-gravity}, we will take the ``quantize first, constrain second'' approach here: first constructing the Hilbert space ignoring $\text{Map}(\Sigma,n)$ and projecting onto invariant wavefunctions at the end.

At fixed particle number $\mathcal M_{\text{grav},n}$ is K\"ahler with symplectic form
\beq\label{eq:cone-symplectic-form}
    \omega=\frac{k}{4\pi}\omega_{\text{WP},n,\theta}-\frac{k}{4\pi}\overline\omega_{\text{WP},n,\theta}~,
\eeq
where
\beq
    \omega_{\text{WP},n,\theta}=4\pi\left(\kappa_1-\sum_{i=1}^{n}\alpha_\theta^2\,\psi_i\right)~,
\eeq
is the Weil-Petersson form on the Teichm\"uller space of cone surfaces.\footnote{The Weil-Petersson form on $\mathcal{T}_{\Sigma,n,\theta}$ plays a central role in the study of Jackiw–Teitelboim (JT) gravity with conical defects \cite{Mertens:2019tcm,Witten:2020wvy,Turiaci:2020fjj,Maxfield:2020ale,Eberhardt:2023rzz,Kruthoff:2024gxc}. See also \cite{Do:2006xyz,Anagnostou:2023xyz} for treatments in the math literature.} The definitions of the characteristic classes $\kappa_1$ and $\psi_i$ are reviewed in Appendix \ref{app:bundles}.
Thus unlike considering the scalar field theory {\it in toto}, we can proceed by geometric quantization at each particle number as an intermediate step. However, before doing so, let us address an important subtlety that arises in the defect picture.

\subsection{Compactification of the phase space}

Before quantizing the $n$-particle phase space, we need to decide how wavefunctions should behave near the boundaries of phase space. In particular, we need a prescription for treating the coincident limit of the point particles on the initial time surface. This amounts to picking a particular compactification of the Teichm\"uller space $\mathcal{T}_{\Sigma,n,\theta}$.\footnote{It is an unfortunate quirk of terminology that the `compactification' of the Teichm\"uller space $\mathcal{T}_{\Sigma,n,\theta}$ is not, in fact, a compact space.}

\begin{figure}
\centering
\begin{tikzpicture}[scale = 0.75]
\begin{scope}
\begin{scope}
\draw[very thick] (0,0) [partial ellipse = 0:360:3 and 0.75];
\draw[very thick] (-1.7,0) to[out = 0, in = -135] (-1,0.4) to[out = -45, in = 180] (-0.3,0);
\draw[fill = CornflowerBlue] (-1,0.4) circle (0.075);
\draw[very thick] (0.3,0) to[out = 0, in = -135] (1,0.4) to[out = -45, in = 180] (1.7,0);
\draw[fill = CornflowerBlue] (1,0.4) circle (0.075);
\draw[very thick, -latex] (3.5,0) -- (6.5,0);
\node[above] at (5,0) {$\sum\theta_i\leq 2\pi$};
\end{scope}
\begin{scope}[xshift = 10cm]
\draw[very thick] (0,0) [partial ellipse = 89:-269:3 and 0.75];
\draw[very thick] (-0.8,0) to[out = 0, in = -95] (0,1) to[out = -85, in = 180] (0.8,0);
\draw[fill = CornflowerBlue] (0,1) circle (0.1);
\end{scope}
\end{scope}
\begin{scope}[yshift = -6cm]
\begin{scope}
\draw[very thick] (0,0) [partial ellipse = 0:360:3 and 0.75];
\draw[very thick] (-1.7,0) to[out = 0, in = -135] (-1,0.4) to[out = -45, in = 180] (-0.3,0);
\draw[fill = CornflowerBlue] (-1,0.4) circle (0.075);
\draw[very thick] (0.3,0) to[out = 0, in = -135] (1,0.4) to[out = -45, in = 180] (1.7,0);
\draw[fill = CornflowerBlue] (1,0.4) circle (0.075);
\draw[very thick, -latex] (3.5,0) -- (6.5,0);
\node[above] at (5,0) {$\sum\theta_i> 2\pi$};
\end{scope}
\begin{scope}[xshift = 10cm]
\draw[very thick] (0,0) [partial ellipse = 0:360:3 and 0.75];
\draw[very thick] (-0.5,0) to[out = 0, in = -90] (0,0.8) to[out = 90, in = -45] (-1.0607,1.9393);
\draw[very thick] (0.5,0) to[out = 180, in = -90] (0,0.8) to[out = 90, in = -135] (1.0607,1.9393);
\draw[very thick] (0,3) [partial ellipse = -45:225:1.5 and 1.5];
\begin{scope}[xshift = -0.8cm, yshift = 3.4cm, rotate = 45]
\draw[very thick] (-0.5,0) to[out = 0, in = -135] (0,0.4) to[out = -45, in = 180] (0.5,0);
\draw[fill = CornflowerBlue] (0,0.4) circle (0.075);
\end{scope}
\begin{scope}[xshift = 0.8cm, yshift = 3.4cm, rotate = -45]
\draw[very thick] (-0.5,0) to[out = 0, in = -135] (0,0.4) to[out = -45, in = 180] (0.5,0);
\draw[fill = CornflowerBlue] (0,0.4) circle (0.075);
\end{scope}
\end{scope}
\end{scope}
\end{tikzpicture}
\caption{When multiple conical defects collide, they can either combine into a single, sharper defect, or `bubble off' into a nodal sphere.}
\label{fig:compactification}
\end{figure}
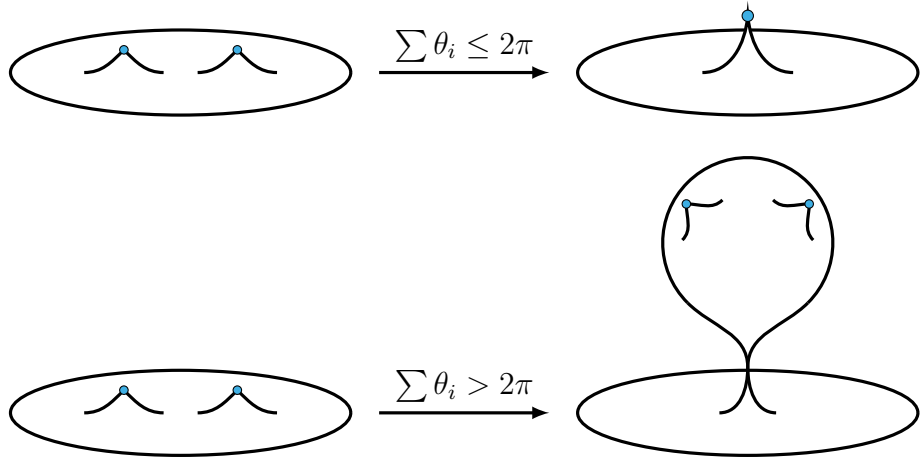

It turns out that there is essentially one choice of compactification that is mathematically natural \cite{Do:2006xyz,Witten:2020wvy,Maxfield:2020ale,Turiaci:2020fjj,Eberhardt:2023rzz,Kruthoff:2024gxc}, and which is depicted in Figure \ref{fig:compactification}. Consider the limit where $p$ points are taken to collide. If the sum of their defect angles is less than $2\pi$, then the result will simply be a sharper conical defect of total defect angle $\theta_{\text{tot}}=p\,\theta$. If, on the other hand, the sum of the defect angles is more than $2\pi$, this is not possible. Instead, in this case, there will be a geodesic that surrounds the $p$ points, and the length of this geodesic can be taken to zero. The resulting surface, shown in the bottom of Figure \ref{fig:compactification}, is a sphere with $p$ conical defects attached to $\Sigma$ by a node.\footnote{More generally, a surface can degenerate by bubbling off many nodal surfaces, provided that each nodal component admits a hyperbolic metric (with conical singularities), just as in the usual Deligne-Mumford compactification of the moduli space of curves \cite{Zvonkine:xyz}.}

Physically, this compactification can be understood as follows. Given $p$ particles of mass $m$, the coincident limit looks like a single particle of mass $pm$, assuming that the total mass does not exceed $1/4G_N$. If the total mass exceeds this threshold, then the classical geometry has a geodesic that encloses all of the particles, and the boundary of the moduli space is found by letting the length of that geodesic shrink to zero. (As we will discuss in more detail in Section \ref{sec:black-holes}, the appearance of the BTZ threshold $1/4G_N$ and the resemblance of the enclosing geodesic to a black hole horizon is not coincidental.)

Below the threshold we can determine the conformal weight of the combined defect from \eqref{eq:conformal-weight-main} as
\begin{equation}
\begin{split}
    h(p)&=k(1-\alpha_{p\theta}^2)\\
    &=k(1-(1-p(1-\alpha_{\theta}))^2)\,,
\end{split}
\end{equation}
or in terms of bulk parameters
\begin{equation}\label{eq:combined_conformal_weight}
\begin{split}
h(p)&=\frac{p\mu}{2}-\frac{G_Np^2\mu^2}{\ell}\,,\\
&=\frac{p}{2}(1+\sqrt{1+m^2\ell^2})-\frac{p^2G_N}{\ell}(1+\sqrt{1+m^2\ell^2})^2\,.
\end{split}
\end{equation}
This formula has an immediate physical interpretation. As we will see later, $2h$ is the energy of a single point particle in global $\text{AdS}_3$ as measured from infinity. As such, we can define the $p$-particle binding energy to be
\begin{equation}
\Delta E_p=2h(p)-2ph=-\binom{p}{2}\frac{4G_N\mu^2}{\ell}\,.
\end{equation}
The minus sign reflects the attractive nature of gravity while the binomial coefficient is precisely what one would expect from the binding energy of a pairwise-interacting system.

\subsection[The ``\texorpdfstring{$n$}{n}-particle'' Hilbert space]{\boldmath The ``\texorpdfstring{$n$}{n}-particle'' Hilbert space}

We now proceed with quantization of $\mc M_{\text{grav},n}$ through the geometric quantization of $\mc T_{\Sigma,n,\theta}$. As before, we will ignore the issue of the mapping class group, $\text{Map}(\Sigma,n)$, instead allowing ourselves the freedom to project our resulting Hilbert space onto invariant wavefunctions at the end.

The geometric quantization of $\mc T_{\Sigma,n,\theta}$ is again given by introducing a pre-quantum line bundle $\mathscr{L}$ over $\mc T_{\Sigma,n,\theta}$, where $\mathscr{L}=\mathcal{L}^k$ with
\begin{equation}
c_1(\mathcal{L})=\frac{1}{4\pi}[\omega_{\text{WP}}]\,.
\end{equation}
Here, $\omega_{\text{WP}}$ is the Weil-Petersson form on $\mc T_{\Sigma,n,\theta}$. A wavefunction, realized as a holomorphic section $\Psi\in\text{H}^0(\mc T_{\Sigma,n,\theta},\mathscr{L})$, depends on the moduli of $\mc T_{\Sigma,n,\theta}$ in precisely the same form as an $n$-point scalar correlation function in a chiral CFT on $\Sigma$ with left-moving central charge $c=24k$:
\beq\label{eq:section_transformation_law}
    \Psi=\Big\langle\mc O_h(x_1)\ldots\mc O_h(x_n)\Big\rangle_{\text{CFT},c}
\eeq
with a conformal weight given by
\beq\label{eq:weight_from_angle}
    h=\frac{c}{24}\left(1-\alpha_\theta^2\right)~.
\eeq
As we explain in Appendix \ref{app:bundles}, the central charge and conformal weights can be read off from the coefficients appearing in the symplectic form \eqref{eq:cone-symplectic-form}. Note that much like the states of the pure gravity Hilbert space, holomorphic sections of $\mathscr{L}$ are conformal blocks for correlators of a putative CFT in the sense that they satisfy the conformal Ward identities, however they do not necessarily satisfy crossing symmetry.

While it is natural and tempting to consider
\beq\label{eq:Hgn0}
    \Hgn^0\equiv\text{H}^0\left(\mc T_{\Sigma,n,\theta},\mathscr{L}\right)\otimes\text{H}^0\left(\overline{\mc T}_{\Sigma,n,\theta},\overline{\mathscr{L}}\right)\,,
\eeq
as an ``$n$-particle'' Hilbert space, we caution that this language is only a heuristic. Even before mapping class group considerations, there is no strict separation between the gravitational and matter degrees of freedom in the above Hilbert space. Physically this is obvious from the fact that in building the phase space as classical solutions, we have allowed the particles to backreact. Moreover, these degrees of freedom are coupled in the definition of $\mc T_{\Sigma,n,\theta}$ through modding by small diffeomorphisms. A more dramatic expression of the coupling lies in the compactification described above: indeed for collisions of numerous enough (or heavy enough) defects, the result is not a defect at all, but instead a drastically backreacted geometry.

\paragraph{The inner product:}

Having established the vector space structure of the ``n-particle" sector, $\mc H_{\text{grav},n}^0$, we now prescribe an inner product for its states. Recall that sections of $\mathscr{L}$ are given by $n$-point conformal blocks of a chiral CFT labelled by $n$ marked points, $(z_1,\ldots, z_n)$. These conformal blocks have central charge $c$ and conformal weight $h$ at the insertions, and so a natural inner product is a given by a generalization of \eqref{eq:vtqft-inner-product}:

\beq\label{eq:npart-Hgm-InnerProd}
    \Braket{ \overline{\Psi}_1,\Psi_2}=\int_{\mc T_{\Sigma,n,\theta}}Z_{bc}\Braket{ e^{2\beta \chi}(z_1)\cdots e^{2\beta\chi}(z_n)}_{\text{tL}}\overline{\Psi}_1(\bar z_1,\ldots,\bar z_n)\Psi_2(z_1,\ldots,z_n)~,
\eeq
where the timelike Liouville momentum $\beta$ is chosen such that conformal weight of each vertex operator $e^{2\beta\chi}$ is $1-h$ and the entire expression is Weyl invariant. Note that the integration over $\mc T_{\Sigma,n,\theta}$ includes an integration over the defect locations $(z_i,\overline z_i)$ as part of the moduli of $\mc T_{\Sigma,n,\theta}$. As it stands, one still has to check that this prescription leads to a positive-definite inner product. This has been explicitly checked for particles of conformal weight $h>\frac{c-1}{24}$ \cite{Collier:2023fwi}, but not for sub-threshold weights corresponding to conical defects.\footnote{We thank Lorenz Eberhardt for discussions on this point.}

\paragraph{Assembling the field theory:}

Having constructed each $\mc H^0_{\text{grav},n}$ we now state our proposal for the gravity and matter Hilbert space as the direct sum over all particle numbers, taking care to project onto mapping class invariant wavefunctions in each sector:
\beq\label{eq:Hgm-total}
    \Hgm=\bigoplus_{n=0}^\infty \Hgn~,\qquad \Hgn\equiv\Hgn^0/\text{Map}(\Sigma,n)~.
\eeq
Note that for each sector in this sum, $\text{Map}(\Sigma,n)$ contains as a subgroup the permutations of the $n$-punctures.\footnote{More correctly it contains as a subgroup the braid group. The permutation group is realized as the braid group quotiented by pure braids which act trivially on our sections.} The inner product \eqref{eq:npart-Hgm-InnerProd} on $\mc H^0_{\text{grav},n}$ is naturally defined on $\mc H_{\text{grav},n}$ by restriction. This then also defines an inner product on $\mc H_\text{grav+matter}$ by the action on each sector.

\paragraph{Metaplectic correction:} An important issue worth mentioning is that, in passing from the classical to quantum descriptions in K\"ahler quantization, an extra step known as \textit{metaplectic correction} is often taken. This has the effect of `correcting' the pre-quantum line bundle $\mathcal{L}^{k}$ by tensoring it with the square root $\mathscr{K}^{1/2}$ of the canonical bundle on $\mathcal{T}_{\Sigma,n,\theta}$. In pure gravity this has the effect of correcting the central charge to $c=24k+13$ and with a minimally coupled scalar also of adding $1/2$ to the conformal weights of the conical defects. In the rest of the paper we will choose not to include this correction, but it may be the case that, when comparing to the path integral of a free scalar field, it is necessary. For the sake of completeness, we include a discussion of the metaplectic correction in Appendix \ref{app:metaplectic_correction}.

\paragraph{An aside on decoupling:}
It is also useful to ask how one recovers standard bosonic quantum field theory on a rigid background from our construction. We can consider the limit of $G_N\rightarrow 0$ while holding the scalar mass $m$ fixed. In this case the backreaction of each particle insertion gives a vanishing conical angle, and so we can take the phase space to consist of standard hyperbolic metrics. It is in this sense that the dynamics of probe quantum particles on a rigid background is recovered in this limit. The promotion of this space to Teichm\"uller space, $\mc T_{\Sigma,n,\theta=0}$, however involves quotienting by small diffeomorphisms which act on the insertion points of the probe particles and so there is no strict sense in which a $G_N\rightarrow 0$ limit of \eqref{eq:Hgm-total} decouples into $\Hg\otimes\mc H_\text{scalar}$.

\section{Partition functions}\label{sec:partition-functions}

Having provided a proposal for the Hilbert space of quantum matter coupled to 3D gravity, we have at our disposal the means to compute some non-trivial observables. One obvious candidate is the gravity and matter path integral over $\Sigma\times S^1$, which we can interpret as a thermal partition function. For compact $\Sigma$, the Hamiltonian vanishes as a constraint and this computes the total Hilbert space dimension on $\Sigma$:
\beq\label{eq:ZasdimH}
    Z_\text{grav+matter}[\Sigma\times S^1]=\text{dim}\,\mathcal H_\text{grav+matter}[\Sigma]~.
\eeq

\begin{figure}
\centering
\begin{tikzpicture}[scale = 1.25]
\draw[very thick, orange, style={decorate, decoration={snake, amplitude = 0.2mm, segment length = 3mm}}] (2,-0.05) [partial ellipse = 0:360:0.4 and 1.9];
\draw[very thick] (2,0) [partial ellipse = 0:360:0.5 and 2];
\draw[very thick] (1.9,1.963) to[out = -125, in = 0] (0,0.25);
\draw[very thick] (1.9,-1.963) to[out = 125, in = 0] (0,-0.25);
\node[left] at (0,0) {$\cdots$};
\node[above] at (0.5,0.5) {$\Sigma$};
\draw[very thick, latex-] (2,0) [partial ellipse = 30:-30:0.7 and 2.2];
\node[right] at (2.8,0) {$q^{L_0}\bar{q}^{\bar{L}_0}$};
\end{tikzpicture}
\caption{A surface with an asymptotic boundary. The boundary modes are depicted as orange wiggles. The Hamiltonian $L_0+\bar{L}_0$ and angular momentum $L_0-\bar{L}_0$ acts as a dilatation and rotation around the boundary circle, respectively.}
\label{fig:trumpet}
\end{figure}
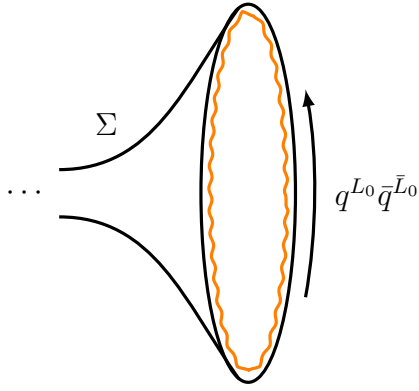

When $\Sigma$ has an asymptotic boundary, we have more structure associated to the pure gravitational phase space. This phase space is infinite dimensional, containing sectors of Alekseev--Shatashvili modes near each asymptotic boundary \cite{Cotler:2018zff}, as depicted in Figure \ref{fig:trumpet}, which are not affected by the inclusion of conical defects. Because of this, the Hilbert space dimension is infinite and equation \eqref{eq:ZasdimH} will diverge.

We can address this issue by modifying the computation in the following way. To each asymptotic boundary we can associate a $\mathbb{C}^*$ action which rotates and dilates the boundary circle and is generated by a boundary Hamiltonian/angular momentum operator. Since the full gravitational phase space (at fixed particle number) is given by two copies of Teichm\"uller space, there are two such Hamiltonians which are identified as the Virasoro generators, $L_0^{(i)}$ and $\bar L_0^{(i)}$, where $i$ labels each boundary. We can then grade the partition function as\footnote{Here we utilize the notation $\bTr{\mc O}{\mc H}$ to indicate the trace of the operator $\mc O$ over the Hilbert space $\mc H$.}
\beq
    Z(\tau_i,\bar\tau_i)=\bTr{\prod_{i}q_i{\vphantom{q_i}}^{L_0^{(i)}}\bar q_i{\vphantom{\bar q_i}}^{\bar L_0^{(i)}}}{\mc H_{\text{grav}+\text{matter}}}~,
\eeq
with
\begin{equation}
q_i=e^{2\pi i\tau_i}\,,\quad \bar{q}_i=e^{-2\pi i\bar\tau_i}\,.
\end{equation}

\subsection{Index theorems}

Let us now specialize to the case of a single asymptotic boundary. Since on a given surface $\Sigma$ with a boundary, the Hilbert space $\Hgm$ is decomposed into multiparticle sectors as in \eqref{eq:Hgm-total} and \eqref{eq:Hgn0}, the partition function admits the decomposition\footnote{For ease of notation, we use the convention that the square of a chiral object is obtained by multiplying that object by its right-moving counterpart.}
\begin{equation}\label{eq:partition-function-decomposition}
Z(\tau,\bar\tau)=\sum_{n=0}^{\infty}\frac{1}{|\text{Map}(\Sigma,n)|}\sum_{g\in\text{Map}(\Sigma,n)}\left|\bTr{gq^{L_0}}{\text{H}^0(\mathcal{T}_{\Sigma,n\theta},\mathscr{L})}\right|^2\,,
\end{equation}
where $\mathscr{L}$ and $\overline{\mathscr{L}}$ are the quantum line bundles on $\mathcal{T}_{\Sigma,n,\theta}$ and $\overline{\mathcal{T}}_{\Sigma,n,\theta}$. The sum over mapping class transformations\footnote{Because $\text{Map}(\Sigma,n)$ has infinitely many elements, this is strictly not a well-defined operation. As we will see below, we can sidestep this issue specifically when $\Sigma$ is the hyperbolic disk, but more generally one has to be careful, for example by taking the trace over the \textit{co-invariant} Hilbert space, see for example \cite{Held:2025mai,Balasubramanian:2025rcr}. We thank Charlie Cummings for discussions on this point.}
\begin{equation}
\frac{1}{|\text{Map}(\Sigma,n)|}\sum_{g\in\text{Map}(\Sigma,n)}g
\end{equation}
projects onto the set of wavefunctions that are crossing-symmetric.

Thus, in order to compute the gravitational partition function on $\Sigma\times\text{S}^1$, we need to compute the traces of $gq^{L_0}$ acting on the space $\text{H}^0(\mathcal{T}_{\Sigma,n,\theta},\mathscr{L})$. This quantity is in general difficult to compute, but as we will briefly explain, the equivariant Euler characteristic
\begin{equation}
\chi(gq^{L_0},\mathscr{L})=\sum_{i=0}^{\infty}(-1)^i\bTr{gq^{L_0}}{\text{H}^i(\mathcal{T}_{\Sigma,n,\theta},\mathscr{L})}
\end{equation}
is computable by an index theorem. Fortunately, the Kodaira vanishing theorem guarantees that for $k$ large enough the higher cohomology groups of the line bundle $\mathscr{L}$ are trivial. Since we are always interested in weakly-coupled gravity, we will take $k\gg 1$, so that it always suffices to compute the Euler characteristic.

The equivariant Euler characteristic can be computed via application of the Atiyah-Bott fixed-point theorem \cite{Atiyah-Bott:1968xxx}. The precise form of the index theorem depends on the form of the fixed-point set of $gq^{L_0}$. Currently, we will only be interested in a case where the fixed-point set is a single point in $\mathcal{T}_{\Sigma,n,\theta}$, so we will simply quote the result in this special case. Let $\star\in\mathcal{T}_{\Sigma,n,\theta}$ be the fixed point of $gq^{L_0}$, let $\mathscr{L}_{\star}$ be the fiber of $\mathscr{L}$ over the fixed point, and let $T_{\star}\mathcal{T}_{\Sigma,n,\theta}$ be the tangent space to the fixed point. Then the equivariant Euler characteristic is given by
\begin{equation}\label{eq:single-fixed-point-index-theorem}
\chi(gq^{L_0},\mathscr{L})=\frac{\bTr{gq^{L_0}}{\mathscr{L}_{\star}}}{\text{det}(1-gq^{L_0}|T_{\star}\mathcal{T}_{\Sigma,n,\theta})}\,.
\end{equation}
Since $\mathscr{L}_{\star}$ is a one-dimensional vector space, the numerator is simply the eigenvalue of $gq^{L_0}$ acting on the fiber. 

\subsection[Thermal AdS\texorpdfstring{$_3$}{3}]{\boldmath Thermal AdS\texorpdfstring{$_3$}{3}}

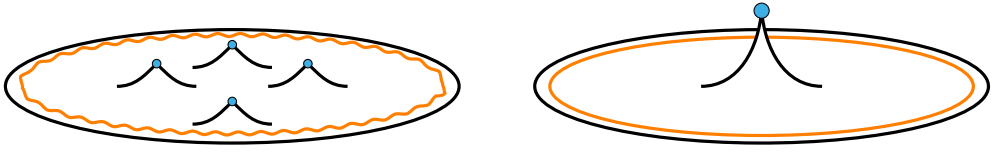
\begin{figure}
\centering
\begin{tikzpicture}
\begin{scope}
\draw[very thick] (0,0) [partial ellipse = 0:360:3 and 0.75];
\begin{scope}[xshift = -1cm, scale = 0.75]
\draw[very thick] (-.7,0) to[out = 0, in = -135] (0,0.4) to[out = -45, in = 180] (.7,0);
\draw[fill = CornflowerBlue] (0,0.4) circle (0.075);
\end{scope}
\begin{scope}[xshift = 1cm, scale = 0.75]
\draw[very thick] (-.7,0) to[out = 0, in = -135] (0,0.4) to[out = -45, in = 180] (.7,0);
\draw[fill = CornflowerBlue] (0,0.4) circle (0.075);
\end{scope}
\begin{scope}[yshift = 0.25cm, scale = 0.75]
\draw[very thick] (-.7,0) to[out = 0, in = -135] (0,0.4) to[out = -45, in = 180] (.7,0);
\draw[fill = CornflowerBlue] (0,0.4) circle (0.075);
\end{scope}
\begin{scope}[yshift = -0.5cm, scale = 0.75]
\draw[very thick] (-.7,0) to[out = 0, in = -135] (0,0.4) to[out = -45, in = 180] (.7,0);
\draw[fill = CornflowerBlue] (0,0.4) circle (0.075);
\end{scope}
\draw[very thick, orange, style={decorate, decoration={snake, amplitude = 0.2mm, segment length = 3mm}}] (0,-0.05) [partial ellipse = 0:360:2.8 and 0.65];
\end{scope}
\begin{scope}[xshift = 7cm]
\draw[very thick] (0,0) [partial ellipse = 89:-269:3 and 0.75];
\draw[very thick, orange] (0,0) [partial ellipse = 89:-269:2.8 and 0.65];
\draw[very thick] (-0.8,0) to[out = 0, in = -95] (0,1) to[out = -85, in = 180] (0.8,0);
\draw[fill = CornflowerBlue] (0,1) circle (0.1);
\end{scope}
\end{tikzpicture}
\caption{\textbf{Left:} A generic point in the Teichm\"uller space $\mathcal{T}_{\mathbb{D},n,\theta}$. \textbf{Right:} The fixed point of $gq^{L_0}$ for which all defects lie at the origin and the boundary modes are constant.}
\label{fig:localization}
\end{figure}

We now arrive at the main application of our formalism: the calculation of the gravitational path integral on thermal $\text{AdS}_3$ coupled to a scalar field. In this case, the initial time surface $\Sigma$ is the hyperbolic disk $\mathbb{D}$, and $\mathcal{T}_{\mathbb{D},n,\theta}$ is the moduli space of hyperbolic metrics on the disk with $n$ conical defects. This is an infinite-dimensional space parametrized by boundary modes near the asymptotic boundary and the locations of the conical defects, see Figure \ref{fig:localization}.

Now, the mapping class group of the disk with $n$ marked points is the braid group $B_n$. The braid group can be roughly decomposed into \textit{pure braids} (those which bring each marked point to their original position) and permutations. More concretely, the braid group sits at the center of a short exact sequence
\begin{equation}
1\to P_n\to B_n\to S_n\to 1\,,
\end{equation}
where $P_n$ is the group of pure braids. Now, let $\Psi$ be a wavefunction in the $n$-particle Hilbert space $\mathcal{H}^0_{\text{grav},n}$. Recalling the interpretation of $\Psi$ as a conformal block for an $n$-point correlator in a chiral CFT, a pure braid acts on $\Psi$ only through a phase. This phase depends on the conformal weights $h$ of the point particles involved in the braids. Since wavefunctions in the full theory (before gauging by the mapping class group) are of the form $\Psi_1\bar{\Psi}_2$, and since all particles have equal left- and right-moving conformal weights, these phases cancel in wavefunctions of the full theory. Thus, we can effectively ignore the effect of pure braids when gauging the mapping class group.

Put another way, we can trade each braid for its underlying permutation $\rho\in S_n$ and consider projecting onto $S_n$-invariant states by trading the sum over all braids for the $S_n$ projector
\begin{equation}
\frac{1}{n!}\sum_{\rho\in S_n}\rho
\end{equation}
in the traces in equation \eqref{eq:partition-function-decomposition}. That is, we compute
\begin{equation}
Z(\tau,\bar\tau)=\sum_{n=0}^{\infty}\frac{1}{n!}\sum_{\rho\in S_n}\left|\bTr{\rho q^{L_0}}{\text{H}^0(\mathcal{T}_{\Sigma,n\theta},\mathscr{L})}\right|^2
\end{equation}
For a given partition, the trace
\begin{equation}
\bTr{\rho q^{L_0}}{\text{H}^0(\mathcal{T}_{\mathbb{D},n,\theta},\mathscr{L})}
\end{equation}
can now be computed by the index theorem \eqref{eq:single-fixed-point-index-theorem}. Specifically, we assume that $n\theta< 2\pi$, so that the compactification of the Teichm\"uller space is simply found by letting the defects fuse. This assumption is actually rather mild, since, as we will discuss in more detail in Section \ref{sec:black-holes}, values of $n$ such that $n\theta\geq 2\pi$ contribute non-perturbatively in $G_N$, so that we can safely ignore them and restrict the sum to $0\leq n\leq\lfloor\ell/4G_N\mu\rfloor$ if we are only interested in perturbative quantities. We will call the partition function under this assumption $Z_\text{pert}$. Since $q^{L_0}$ applies a rotation of the disk around the origin, the transformation $\rho q^{L_0}$ has a single fixed point. On this fixed point,  the conical defects are all taken to lie at the origin and the boundary mode is completely round, see Figure \ref{fig:localization}.

At this fixed-point, the fiber $\mathscr{L}_{\star}$ is invariant under permutations and transforms under boundary rotations like the disk one-point function $\Braket{V_{h(n)}(0)}$ in a chiral CFT of central charge $c$, where $h(n)$ is defined in equation \eqref{eq:combined_conformal_weight}. This one-point function transforms under $q^{L_0}$ like
\begin{equation}
\braket{V_{h(n)}(0)}\to q^{h(n)}q^{-c/24}\Braket{V_{h(n)}(0)}\,,
\end{equation}
where the $q^{h(n)}$ comes from the transformation of a Virasoro primary, while $q^{-c/24}$ comes from the global conformal anomaly on the disk. Thus, the numerator of the index theorem is\footnote{This result could also have been obtained by extending the classes $\kappa_1$ and $\psi_i$ in equivariant cohomology, see \cite{Eberhardt:2023rzz}.}
\begin{equation}
\bTr{\rho q^{L_0}}{\mathscr{L}_{\star}}=q^{h(n)-c/24}\,.
\end{equation}

The denominator is more complicated. The tangent space $T_{\star}\mathcal{T}_{\mathbb{D},n,\theta}$ is spanned by infinitesimal boundary modes and infinitesimal separations of the locations of the point particles. The boundary modes correspond to descendants of the Virasoro vacuum and thus have eigenvalues $L_0=2,3,\ldots$ under rotation but are unaffected by permutation of the point particles. Thus, they contribute a factor of
\begin{equation}
\prod_{m=2}^{\infty}\frac{1}{1-q^m}\,.
\end{equation}
Next, the part of $T_{\star}\mathcal{T}_{\mathbb{D},n,\theta}$ spanned by infinitesimal separations of the conical defects are parametrized by an $n$-tuple of coordinates $(z_1,\ldots,z_n)$, and $\rho q^{L_0}$ acts on these coordinates as
\begin{equation}
(z_1,\ldots,z_n)\to(q z_{\rho(1)},\ldots,qz_{\rho(n)})\,.
\end{equation}
As such, the eigenvalues of $\rho q^{L_0}$ are simply the eigenvalues of $\rho$ multiplied by $q$. Any $\rho\in S_n$ can be decomposed into a disjoint set of cycles. For a given cycle length $r$, let $\ell_r$ be the number of cycles in $\rho$ of that length. Then the eigenvalues of $\rho$ are the $r^{\text{th}}$ roots of unity with multiplicity $\ell_r$. With this in mind, the separation modes contribute
\begin{equation}
\prod_{r=1}^{n}\prod_{s=0}^{r-1}\frac{1}{(1-qe^{2\pi i s/r})^{\ell_r}}=\prod_{r=1}^{n}\frac{1}{(1-q^{r})^{\ell_r}}\,.
\end{equation}
to the denominator of the index theorem \eqref{eq:single-fixed-point-index-theorem}.

Thus, we have arrived at the value
\begin{equation}
\bTr{\rho q^{L_0}}{\text{H}^{0}(\mathcal{T}_{\mathbb{D},n,\theta})}=q^{h(n)-c/24}\prod_{m=2}^{\infty}\frac{1}{(1-q^m)}\prod_{r=1}^{n}\frac{1}{(1-q^r)^{\ell_r}}\,.
\end{equation}
Adding in the right-moving sector, we are led to the partition function
\begin{equation}\label{eq:thermal-ads3-partition-function}
Z_{\text{pert}}(\tau,\bar\tau)= |q|^{-c/12}\prod_{m=2}^{\infty}\frac{1}{|1-q^m|^2}\sum_{n=0}^{\lfloor \ell/4G_N\mu\rfloor}|q|^{2h(n)}\prod_{r=1}^{n}\sum_{\{\ell_1,\ldots,\ell_n\}}\frac{1}{\ell_r!r^{\ell_r}}\frac{1}{|1-q^r|^{2\ell_r}}\,,
\end{equation}
Note that the combinatorial factors in equation \eqref{eq:thermal-ads3-partition-function} arise by taking into account the number of distinct permutations with a given cycle type, and the internal sum is over all partitions $\{\ell_1,\ldots,\ell_n\}$ of $n$ (i.e. $\sum_{r=1}^{n}r\ell_r=n$).

\subsection{A three-dimensional picture: the Wilson spool}\label{sec:spool}

Equation \eqref{eq:thermal-ads3-partition-function} is the main technical result of this paper. In the next two subsections, we will put it to a basic sanity test and explore its implications for higher-loop diagrammatics in thermal $\text{AdS}_3$. First, however, let us take a moment to interpret the terms appearing.

\begin{figure}
\centering
\begin{tikzpicture}[scale = 1.5]
\draw[very thick] (0,0) [partial ellipse = 0:-180:1.5 and 0.375];
\draw[very thick, dashed] (0,0) [partial ellipse = 0:180:1.5 and 0.375];
\draw[line width = 3.5, white] (0,0.1) to[out = 90, in = -90] (-0.75,2.9);
\draw[very thick, MidnightBlue] (0,0.1) to[out = 90, in = -90] (-0.75,2.9);
\draw[line width = 3.5, white] (-0.75,-0.1) to[out = 90, in = -90] (0.75,2.9);
\draw[very thick, MidnightBlue] (-0.75,-0.1) to[out = 90, in = -90] (0.75,2.9);
\draw[line width = 3.5, white] (0.75,-0.1) to[out = 90, in = -90] (0,3.1);
\draw[very thick, MidnightBlue] (0.75,-0.1) to[out = 90, in = -90] (0,3.1);
\draw[line width=3pt, white] (0,3) [partial ellipse = 0:360:1.5 and 0.375];
\draw[very thick] (0,3) [partial ellipse = 0:360:1.5 and 0.375];
\draw[very thick] (-1.5,0) -- (-1.5,3);
\draw[very thick] (1.5,0) -- (1.5,3);
\end{tikzpicture}
\caption{A three-dimensional picture of the trace of $gq^{L_0}\bar{q}^{\bar{L}_0}$ as a Euclidean path integral of braided worldlines.}
\label{fig:braided-loops}
\end{figure}
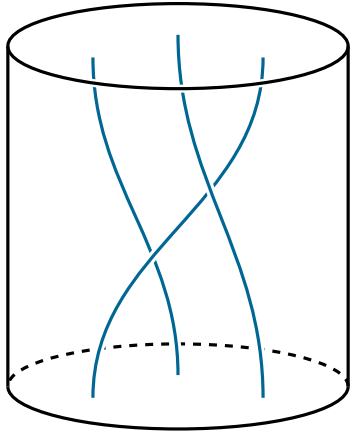

In a three-dimensional picture, each point particle in $\mathbb{D}\times\text{S}^1$ sweeps out a worldline that winds the thermal circle. Let $g$ again be an element of the mapping class group of the $n$-punctured disk, i.e. a braid. Then the trace
\begin{equation}\label{eq:braid-trace}
\bTr{g q^{L_0}\bar{q}^{\bar{L}_0}}{\mathcal{H}^0_{\text{grav}+\text{matter}}}
\end{equation}
appearing in the partition function is computed by the Euclidean path integral in Figure \ref{fig:braided-loops}, where the initial and final time surfaces are identified. The sum over mapping class group elements then has a very natural geometric interpretation: one sums over all possible braidings of particle worldlines in the bulk.

In the Chern-Simons perspective, point particle worldlines are Wilson lines in a given representation $R$ of $\text{SL}(2,\mathbb{R})_L\times\text{SL}(2,\mathbb{R})_R$ determined by the mass of the particle \cite{Witten:1989sx,Carlip:1989nz,Ammon:2013hba}.\footnote{The more rigorous statement in gravity is that point particle worldlines are Wilson lines in the Virasoro TQFT \cite{Collier:2023fwi} whose conformal weights are below the Liouville threshold $(c-1)/24$.} For a given braid $g$, we can associate a path $\gamma(g)$ in thermal $\text{AdS}_3$, where the number of connected components of $\gamma(g)$ is the number of cycles of the braid. The trace \eqref{eq:braid-trace} is interpreted in Chern-Simons language as the expectation value of the Wilson line
\begin{equation}
\mathcal{W}_g=\text{Tr}_R\left[\mathcal{P}\exp\left(\int_{\gamma(g)}(A_L-A_R)\right)\right]\,.
\end{equation}
Because of the relative minus sign between $A_L$ and $A_R$, the expectation value of $\mathcal{W}_g$ does not depend on the precise braiding of the lines, but rather only on the underlying permutation. As such, each term appearing in equation \eqref{eq:thermal-ads3-partition-function} should be interpreted as the expectation value of the Wilson line $\mathcal{W}_{\rho}$ associated to the permutation $\rho$.

Interestingly, the sum over all permutations can be given a particularly simple and suggestive form. If we denote by $\mathcal{W}^{(n)}$ the connected Wilson line that winds the thermal circle $n$ times, then for a permutation $\rho$, we have
\begin{equation}
\mathcal{W}_{\rho}=\prod_{r=1}^{n}(\mathcal{W}^{(r)})^{\ell_r}\,.
\end{equation}
Summing over all permutations and particle number, we are led to the conclusion that the partition function on thermal $\text{AdS}_3$ is given by
\begin{equation}\label{eq:partition-to-spool}
Z(\tau,\bar\tau)=\sum_{n=0}^{\infty}\sum_{\{\ell_1,\ldots,\ell_n\}}\Braket{\prod_{r=1}^{n}\frac{1}{\ell_r!r^{\ell_r}}(\mathcal{W}^{(r)})^{\ell_r}}_{\text{grav}}=\Big\langle\exp\mathbb{W}\Big\rangle_{\text{grav}}\,.
\end{equation}
where
\begin{equation}
\mathbb{W}=\sum_{k=1}^{\infty}\frac{1}{k}\mathcal{W}^{(k)}
\end{equation}
is the \textit{Wilson spool} in anti-de Sitter space. This object was originally defined in \cite{Castro:2023dxp,Castro:2023bvo} as a proposed prescription for coupling matter to 3D gravity in the Chern-Simons prescription. Our analysis here naturally reproduces this proposal from a canonically quantized perspective. We regard our expression \eqref{eq:thermal-ads3-partition-function} as a conjecture for the expectation value of the Wilson spool in 3D gravity perturbatively in $G_N$.

\subsection{Comparison to Giombi-Maloney-Yin}\label{sec:comparison}

The partition function in equation \eqref{eq:thermal-ads3-partition-function} constitutes a conjectural formula for the path integral of $\text{AdS}_3$ gravity coupled to a massive scalar field in a thermal background, valid perturbatively to all orders in $G_N$. As a sanity check of this answer, let us show that the leading-order term reproduces the known one-loop partition functions computed by Giombi, Maloney, and Yin (GMY) \cite{Giombi:2008vd}.

Firstly, if we are interested only in the perturbative behavior of the partition function as a function of $G_N$, we can consistently take $\lfloor1/G_N\mu\rfloor\to\infty$ in the sum over $n$. Once we do this, the only terms in \eqref{eq:thermal-ads3-partition-function} that depend on the gravitational coupling are the ground-state-energy $|q|^{-c/12}$ and the $n$-particle energy $|q|^{2h(n)}$. The former is the value of the renormalized tree-level Einstein-Hilbert action on thermal $\text{AdS}_3$. From the relation \eqref{eq:combined_conformal_weight}, we can write
\begin{equation}
2h(n)=n\mu-\frac{2n^2G_N\mu^2}{\ell}\,.
\end{equation}
Thus, we can expand the partition function in $G_N$ by expanding $|q|^{2h(n)}$. To leading order, it suffices to make the replacement $|q|^{2h(n)}\to |q|^{n\mu}$. At this order in perturbation theory, the sum over partitions of $n$ can be performed explicitly, and we find
\begin{equation}
\begin{split}
Z_{\text{pert}}(\tau,\bar\tau)&=|q|^{-c/12}\prod_{m=2}^{\infty}\frac{1}{|1-q^m|^{2}}\left(\sum_{n=0}^{\infty}\prod_{r=1}^{n}\sum_{\{\ell_1,\ldots,\ell_n\}}\frac{|q|^{\mu r\ell_r}}{\ell_r!r^{\ell_r}|1-q^r|^{2\ell_r}}+\mathcal{O}(G_N\ell^{-1} \mu^2)\right)\\
&=Z_\text{grav,1-loop}\times Z_\text{scalar,1-loop}\times\bigg(1+\mathcal{O}\left(G_N\ell^{-1}\mu^2\right)\bigg)\,.
\end{split}
\end{equation}
with
\beq\label{eq:GMY1loops}
\begin{split}
Z_\text{grav,1-loop}=&|q|^{-c/12}\prod_{m=2}^{\infty}\frac{1}{|1-q^m|^2}~,\\
Z_\text{scalar,1-loop}=&\prod_{l,\bar l=0}^{\infty}\frac{1}{(1-q^{\mu/2+l}\,\bar{q}^{\mu/2+\bar l})}~.
\end{split}
\eeq
$Z_\text{grav,1-loop}$ is the usual Maloney-Witten one-loop determinant of $\text{AdS}_3$ gravity \cite{Maloney:2007ud}, while $Z_\text{scalar,1-loop}$ is precisely the one-loop determinant of a massive scalar field in $\text{AdS}_3$ \cite{Giombi:2008vd}, provided we make the identification \eqref{eq:self_energy_mass_relation} between the point-particle coupling $\mu$ and the scalar mass $m$.

\subsection{Extracting higher-loop terms}\label{sec:higher-order}

Our formalism also allows for an efficient method to calculate higher-loop (in $G_N$ perturbation theory) corrections to the GMY result. For simplicity let us take $q=\bar q=e^{-\beta/\ell}$ and noting
\begin{equation}
|q|^{2h(n)}=e^{-\frac{\beta}{\ell}\left(n\mu-n^2\left(\frac{2G_N}{\ell}\right)\mu^2\right)}=e^{-\frac{n\beta\mu}{\ell}}\sum_{s=0}^{\infty}\frac{1}{s!}\left(\frac{2n^2G_N\beta \mu^2}{\ell^{2}}\right)^s\,,
\end{equation}
We can read off the $(s+1)$-loop contribution to the thermal AdS$_3$ partition function from the $(G_N)^s$ coefficient of the Taylor expansion of \eqref{eq:thermal-ads3-partition-function}:
\begin{equation}
\begin{split}
\frac{Z_{\text{pert}}(\beta)}{Z_{\text{grav,1-loop}}(\beta)}=Z_{\text{scalar,1-loop}}+\sum_{s=1}^{\infty}\frac{(2G_N\beta\mu^2)^s}{\ell^{2s}s!}\sum_{n=0}^{\infty}\sum_{\{\ell_1,\ldots,\ell_n\}}\prod_{r=1}^{n}\frac{1}{l_r!r^{l_r}}\frac{n^{2s}e^{-\frac{n\beta\mu}{\ell}}}{(1-e^{-r\beta/\ell})^{2\ell_r}}\,.
\end{split}
\end{equation}

The sums for $s\geq 1$ seem very difficult to compute. However, note that we can always pull down the $n^{2s}$ factor by starting with the $s=0$ answer (i.e. the 1-loop determinant) and taking $2s$ derivatives with respect to $\mu$. Since $Z_\text{grav,1-loop}$ does not depend on the scalar mass, this gives us the expression
\begin{equation}\label{eq:loop-taylor-series}
Z_{\text{pert}}(\beta)=\sum_{s=0}^{\infty}\frac{1}{s!}\left(\frac{2\mu^2G_N}{\beta}\right)^s\frac{\partial^{2s}}{\partial \mu^{2s}} Z_\text{1-loop}(\mu)\,,
\end{equation}
where $Z_\text{1-loop}(\mu)=Z_\text{grav,1-loop}\times Z_\text{scalar,1-loop}(\mu)$. This provides an extremely efficient method for extracting the higher-loop terms. For instance we report on the two-loop correction as\footnote{We denote by $f|_{G_N^s}$ the coefficient of $G_N^s$ in a perturbative expansion of $f$.}
\begin{equation}\label{eq:two-loop-prediction}
    \frac{\left.Z_{\text{pert}}(\beta)\right|_{G_N}}{Z_\text{1-loop}}=\frac{2G_N\beta\mu^2}{\ell^2}\left[\sum_{l,\bar l=0}^\infty\sum_{r=1}^\infty r\,e^{-\frac{\beta}{\ell}r(\mu+l+\bar l)}+\bigg(\sum_{l,\bar l=0}^\infty\sum_{r=1}^\infty e^{-\frac{\beta}{\ell}r(\mu+l+\bar l)}\bigg)^2\right]
\end{equation}
The sums over $(l,\bar l)$ can be expressed in a compact form as $\text{SL}(2,\mathbb R)$ characters with highest weight $h$,
\begin{equation}
    \frac{\left.Z_{\text{pert}}(\beta)\right|_{G_N}}{Z_\text{1-loop}}=\frac{2G_N\mu^2}{\beta}\left[\left(\sum_{r=1}^{\infty}\chi_h\left(e^{-\frac{\beta}{\ell}r}\right)^2\right)^2-\frac{\ell}{\beta}\partial_\mu\left(\sum_{r=1}^\infty \chi_h\left(e^{-\frac{\beta}{\ell}r}\right)^2\right)\right]
\end{equation}
where
\beq
    \chi_h\left(e^{-\alpha}\right)=\frac{e^{-\alpha h}}{(1-e^{-\alpha})}~.
\eeq
The appearance of these characters is not a surprise: this is a manifestation of $Z_\text{scalar,1-loop}$ expressed as a Wilson spool with an on-shell holonomy in thermal AdS$_3$ equal to $\beta/\ell$ \cite{Castro:2023bvo}. Here the sum over $r$ is the winding of that holonomy around the thermal cycle.

\section{Nonperturbative effects from resummation}\label{sec:black-holes}

The partition function \eqref{eq:thermal-ads3-partition-function} we have derived is perturbative, but it is exact to all orders in $G_N$ perturbation theory. We can formally resum this perturbation theory in the following way. We note that \eqref{eq:loop-taylor-series} implies the differential equation
\beq\label{eq:heateq}
    \partial_{G_N}\left(\frac{Z_{\text{pert}}(\beta)}{Z_\text{grav,1-loop}}\right)=\frac{2\mu^2}{\beta}\partial_\mu^2\left(\frac{Z_{\text{pert}}(\beta)}{Z_\text{grav,1-loop}}\right)~,
\eeq
whose solution admits the closed form integral expression\footnote{This follows from noting that \eqref{eq:heateq} can be mapped to a heat equation. The integral kernel is simply the heat kernel with the initial condition that $\displaystyle\lim_{G_N\rightarrow 0}Z_\text{pert}(\beta)=Z_\text{1-loop}$.},
\beq\label{eq:BMF}
    Z_{\text{pert}}(\beta)=\sqrt{\frac{\beta}{8\pi G_N\mu^2}}\int_{-\infty}^{\infty}\mathrm{d}\bar\mu\,e^{-\beta(\mu-\bar\mu)^2/(8G_N\mu^2)}Z_{\text{1-loop}}(\bar\mu)\,.
\eeq
Within this answer is the cumulative sum of (off-shell) graviton and scalar Feynman diagramatics which, surprisingly, can be captured by integrating the semiclassical one-loop partition function against a Gaussian kernel.

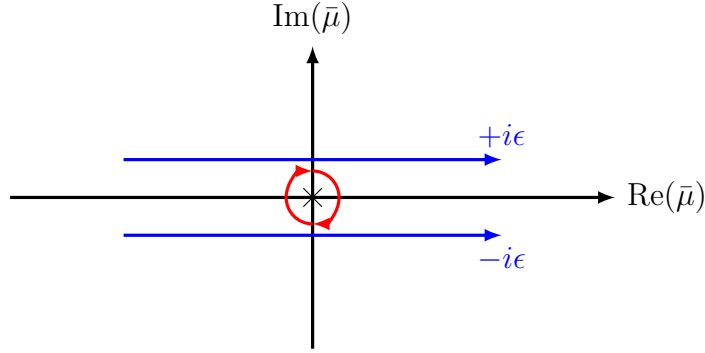
\begin{figure}[h!]
    \centering
    \begin{tikzpicture}[scale=2]
    \draw[very thick, -latex] (-2, 0) -- (2, 0) node[right] {$\text{Re}(\bar\mu)$};
    \draw[very thick, -latex] (0, -1) -- (0, 1) node[above] {$\text{Im}(\bar\mu)$};

    \draw[very thick, -latex,blue] (-1.25,0.25) -- (1.25,0.25) node[above] {$+i\epsilon$};
    \draw[very thick, -latex,blue] (-1.25,-0.25) -- (1.25,-0.25) node[below] {$-i\epsilon$};
    \draw[very thick, latex-,red] (0,.175) arc[start angle = 90, end angle = 270, radius = .175];
    \draw[very thick, latex-,red] (0,-.175) arc[start angle = 270, end angle = 450, radius = .175];

    \node[cross out, draw=black, inner sep=0pt, outer sep=0pt, minimum size=7] at (0, 0) {};
\end{tikzpicture}
    \caption{The ambiguity in the $i\epsilon$ prescription for avoiding poles along the real $\bar\Delta$ axis (depicted in blue) leads to an ambiguity in regulating divergences in the perturbative resummation. The contour wrapping this pole (depicted in red) tells us the free energy of the state needed to resolve this ambiguity.}\label{fig:mucont}
\end{figure}

Additionally surprising is that this all-loop expression for the partition function also encodes certain nonperturbative effects that we have neglected by regarding $G_N$ as a formal infinitesimal parameter. To see this we note that in passing from \eqref{eq:loop-taylor-series} to \eqref{eq:BMF} we have implicitly assumed the integration contour of $\bar\mu$ does not encounter any poles. When $Z_\text{1-loop}(\bar\mu)$ possesses a pole along the real axis, this signals a divergence in the perturbation theory. The standard expectation is that nonperturbative effects come into play in resolving this divergence. In a more pragmatic approach we can tame any divergence associated to real $\bar\mu$ poles by prescribing an $i\epsilon$ prescription for deforming the $\bar\mu$ contour to avoid the pole. The sign of this $i\epsilon$ prescription presents a genuine ambiguity in regulating \eqref{eq:BMF} which, again, one presumes would be resolved by the inclusion of nonperturbative states. We can determine the free energy of such states by looking at the residue of the contour wrapping this pole; see Figure \ref{fig:mucont} for a cartoon.

From \eqref{eq:GMY1loops} we see that the first pole encountered along the $\bar\mu$ axis lies at $\bar\mu=0$. This is a simple pole and its residue yields
\begin{equation}\label{eq:BTZpole}
\begin{split}
   \underset{\bar\mu=0}{\text{Res}}\left[\frac{Z_{\text{pert}}(\beta)}{Z_{\text{grav,1-loop}}(\beta)}\right]&=\sqrt{\frac{\beta}{8\pi G_N\mu^2}}\frac{\ell}{\beta} e^{-\frac{\beta}{8G_N}}\sideset{}{'}\prod_{l,\bar{l}=0}^{\infty}\left(1-e^{-\frac{\beta}{\ell}(l+\bar l)}\right)^{-1}\,,
\end{split}
\end{equation}
where the primed product means that the divergent term $l=\bar{l}=0$ is removed. The exponential behavior signals that this is a potential nonperturbative correction to the integral \eqref{eq:BMF}. Writing the exponential term in terms of the central charge, we find that this residue contributes a term of order $e^{-\frac{\beta}{8G_N}}=|q|^{c/12}$. Thus our resummation of perturbation theory is telling us that the first nonperturbative correction appears precisely at the BTZ threshold!

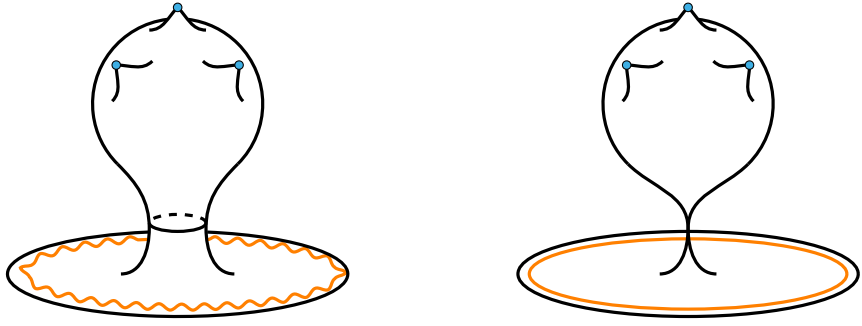
\begin{figure}[h!]
\centering
\begin{tikzpicture}[scale = 0.75]
\begin{scope}
\draw[very thick, orange, style={decorate, decoration={snake, amplitude = 0.4mm, segment length = 3mm}}] (0,0) [partial ellipse = 79:-259:2.8 and 0.62];
\draw[very thick] (0,0) [partial ellipse = 80:-260:3 and 0.75];
\draw[very thick] (0,0.9) [partial ellipse = 0:-180:0.5 and 0.15];
\draw[very thick, dashed] (0,0.9) [partial ellipse = 0:180:0.5 and 0.15];
\draw[very thick] (-1,0) to[out = 0, in = -90] (-0.5,0.8) to[out = 90, in = -45] (-1.0607,1.9393);
\draw[very thick] (1,0) to[out = 180, in = -90] (0.5,0.8) to[out = 90, in = -135] (1.0607,1.9393);
\draw[very thick] (0,3) [partial ellipse = -45:83:1.5 and 1.5];
\draw[very thick] (0,3) [partial ellipse = 97:225:1.5 and 1.5];
\begin{scope}[xshift = -0.8cm, yshift = 3.4cm, rotate = 45]
\draw[very thick] (-0.5,0) to[out = 0, in = -135] (0,0.4) to[out = -45, in = 180] (0.5,0);
\draw[fill = CornflowerBlue] (0,0.4) circle (0.075);
\end{scope}
\begin{scope}[xshift = 0.8cm, yshift = 3.4cm, rotate = -45]
\draw[very thick] (-0.5,0) to[out = 0, in = -135] (0,0.4) to[out = -45, in = 180] (0.5,0);
\draw[fill = CornflowerBlue] (0,0.4) circle (0.075);
\end{scope}
\begin{scope}[yshift = 4.3cm]
\draw[very thick] (-0.5,0) to[out = 0, in = -135] (0,0.4) to[out = -45, in = 180] (0.5,0);
\draw[fill = CornflowerBlue] (0,0.4) circle (0.075);
\end{scope}
\end{scope}
\begin{scope}[xshift = 9cm]
\draw[very thick, orange] (0,0) [partial ellipse = 0:360:2.8 and 0.62];
\draw[very thick] (0,0) [partial ellipse = 0:360:3 and 0.75];
\draw[very thick] (-0.5,0) to[out = 0, in = -90] (0,0.8) to[out = 90, in = -45] (-1.0607,1.9393);
\draw[very thick] (0.5,0) to[out = 180, in = -90] (0,0.8) to[out = 90, in = -135] (1.0607,1.9393);
\draw[very thick] (0,3) [partial ellipse = -45:83:1.5 and 1.5];
\draw[very thick] (0,3) [partial ellipse = 97:225:1.5 and 1.5];
\begin{scope}[xshift = -0.8cm, yshift = 3.4cm, rotate = 45]
\draw[very thick] (-0.5,0) to[out = 0, in = -135] (0,0.4) to[out = -45, in = 180] (0.5,0);
\draw[fill = CornflowerBlue] (0,0.4) circle (0.075);
\end{scope}
\begin{scope}[xshift = 0.8cm, yshift = 3.4cm, rotate = -45]
\draw[very thick] (-0.5,0) to[out = 0, in = -135] (0,0.4) to[out = -45, in = 180] (0.5,0);
\draw[fill = CornflowerBlue] (0,0.4) circle (0.075);
\end{scope}
\begin{scope}[yshift = 4.3cm]
\draw[very thick] (-0.5,0) to[out = 0, in = -135] (0,0.4) to[out = -45, in = 180] (0.5,0);
\draw[fill = CornflowerBlue] (0,0.4) circle (0.075);
\end{scope}
\end{scope}
\end{tikzpicture}
\caption{\textbf{Left:} A generic point in $\mathcal{T}_{\mathbb{D},n,\theta}$ when $n> n_{\text{critical}}$. The conical defects can all lie behind a geodesic on the initial time surface. \textbf{Right:} The fixed-point set of $q^{L_0}$ is taken by shrinking the geodesic to zero size, and is a copy of $\mathcal{T}_{0,\{\theta,\ldots,\theta,2\pi\}}$.}
\label{fig:bh-fixed-point}
\end{figure}

This has a very natural interpretation in terms of the different compactification behaviors shown in Figure \ref{fig:compactification}. The number of particles $n$ for which the compactification transitions is
\begin{equation}
n_{\text{critical}}=\left\lfloor\frac{2\pi}{\theta}\right\rfloor=\left\lfloor\frac{\ell}{4G_N\mu}\right\rfloor\,.
\end{equation}
The conformal weight of the fused defects at this threshold is precisely $c/24$, as can be seen from \eqref{eq:combined_conformal_weight}. For particle numbers $n>n_{\text{critical}}$, the calculation of the partition function becomes more complicated than for small $n$. The trace
\begin{equation}
\bTr{gq^{L_0}}{\text{H}^0(\mathcal{T}_{\mathbb{D},n,\theta},\mathscr{L})}
\end{equation}
is still computed by an equivariant index theorem, however the fixed-point set is no longer a single point. Rather, the conical defects can now live behind a geodesic boundary on the disk, and the fixed-point set occurs when the length of that geodesic is taken to zero, see Figure \ref{fig:bh-fixed-point}. Since a cusp can be thought of as a conical defect with angle $2\pi$ the fixed-point set is a copy of the Teichm\"uller space
\begin{equation}
\Big(\mathcal{T}_{0,\{\theta,\ldots,\theta,2\pi\}}\Big)_g
\end{equation}
inside of $\mathcal{T}_{\mathbb{D},n,\theta}$. The subscript means that we restrict to the fixed-point set of the braid $g$, which corresponds to taking some of the defects within the nodal sphere to lie on top of each other. The generalization of the equivariant index theorem in this case reduces to the computation of the trace only on the Hilbert space associated to the fixed-point set \cite{Eberhardt:2022wlc}
\begin{equation}\label{eq:bh-fixed-point-theorem}
\begin{split}
&\bTr{gq^{L_0}}{\text{H}^0(\mathcal{T}_{\mathbb{D},n,\theta},\mathscr{L})}\\
&\hspace{2cm}=\prod_{m=1}^{\infty}(1-q^m)^{-1}\sum_{r=0}^{\infty}q^r\bTr{g}{\text{H}^0(\mathbb{L}_{n+1}^{r}\otimes\mathscr{L},\mathcal{T}_{0,\{\theta,\ldots,\theta,2\pi\}})}\,,
\end{split}
\end{equation}
where $\mathbb{L}_{n+1}$ is the line bundle whose first Chern class is the $\psi$-class at the node on the bubbled-off sphere in Figure \ref{fig:bh-fixed-point}. The pre-factor intuitively accounts for Virasoro descendants of the primary operators counted in the trace on the right-hand-side. We indeed see that the first term in the $q$-expansion is always proportional to $q^0$ regardless of the value of $n$, verifying that the resulting states all live above the black hole threshold.

How many states, then, live above the black hole threshold? Let's work out the contribution from $g=\text{id}$ (or equivalently any pure braid). In this case the trace simply computes the number of holomorphic sections of $\mathbb{L}_{n+1}^{r}\otimes\mathscr{L}$, which can be computed by the Hirzebruch-Riemann-Roch theorem:
\begin{equation}
\text{dim}\,\text{H}^0(\mathbb{L}_{n+1}^{r}\otimes\mathscr{L},\mathcal{T}_{0,\{\theta,\ldots,\theta,2\pi\}})=\int_{\mathcal{T}_{0,\{\theta,\ldots,\theta,2\pi\}}}\text{td}(T\mathcal{T}_{0,\{\theta,\ldots,\theta,2\pi\}})e^{c_1(\mathscr{L})+r\psi_{n+1}}\,.
\end{equation}
Since $c_1(\mathscr{L})=\frac{k}{4\pi}[\omega_{\text{WP}}]$, the large-$k$ (semiclassical) limit of this integral is dominated by the term
\begin{equation}
\left(\frac{k}{4\pi}\right)^{n-2}\int_{\mathcal{T}_{0,\{\theta,\ldots,\theta,2\pi\}}}\frac{[\omega_{\text{WP}}]^{n-2}}{(n-2)!}\,,
\end{equation}
which is simply the volume of $\mathcal{T}_{0,\{\theta,\ldots,\theta,2\pi\}}$. Since $\mathcal{T}_{0,\{\theta,\ldots,\theta,2\pi\}}$ is non-compact, it has infinite volume, and we conclude that
\begin{equation}
\text{dim}\,\text{H}^0(\mathbb{L}_{n+1}^{r}\otimes\mathscr{L},\mathcal{T}_{0,\{\theta,\ldots,\theta,2\pi\}})=\infty\,.
\end{equation}
Thus we are led to the strange conclusion that the degeneracy of states above the black hole threshold is infinite.

How can we interpret this result? It is tempting to think of the geodesic in Figure \ref{fig:bh-fixed-point} as a Euclidean analogue of a black hole horizon. For this interpretation to be correct, we would expect that the number of states at a given energy above the black hole threshold should reproduce the Bekenstein-Hawking entropy of the black hole. That is, a state of energy $h+\bar{h}=\Delta$ should have degeneracy
\begin{equation}\label{eq:bh-entropy}
d(\Delta)\sim e^{S_{\text{BH}}(\Delta)}=e^{4\pi\sqrt{k\Delta}}\,.
\end{equation}
What our calculation shows is that, while the number of states above the black hole threshold \textit{does} drastically increase, it does so in an uncontrolled way that does not, at least on first glance, resemble the expected result from the black hole interpretation. Instead we find a divergent degeneracy $d(\Delta)=\infty$.

One interpretation of this result is that the theory of gravity coupled to a free scalar is simply sick at energy scales above $\Delta=c/12$. The infinite degeneracy would then be a signifier that coupling a free scalar field to gravity ruins the UV-finiteness of the theory. This is similar to the story in JT gravity: the pure gravity theory is UV finite and described by a random matrix integral \cite{Saad:2019lba} however the inclusion of a scalar displays UV divergences (albeit in a different manner than above) \cite{Jafferis:2022wez}. This is perhaps unsurprising -- the number of field theories coupled to gravity that can successfully be UV completed is expected to be very small. On the other hand, the gravitational path integral is more than just a QFT on a fixed background topology, and it may be possible that the sum over all bulk topologies may be interpreted in such a way as to render the theory UV finite. We will comment more on this question as well as other aspects of the above calculation in the next section.

\section{Discussion}\label{sec:discussion}

In this article we constructed an exact quantization of massive scalar field theory minimally coupled to quantum gravity in three dimensions. Our construction utilizes intuition from the worldline formalism of scalar field theory and the Wilson spool program, as well as technology from the quantization of pure gravity in the Virasoro TQFT. We illustrated how this quantization provides effective methods, namely equivariant index theorems, for computing quantum partition functions of gravity coupled to matter. As an example, we evaluated the partition function of a free scalar field minimally coupled to gravity on thermal AdS$_3$ to all orders in $G_N$ perturbation theory. We explained how this partition function connects to the Wilson spool in thermal AdS$_3$ and showed how it reproduces the scalar one-loop determinant on a fixed thermal AdS$_3$ background at leading order in $G_N$. While this partition function is derived in a formal $G_N\rightarrow 0$ perturbation theory, we showed how the resummation of this perturbation signals the presence of nonperturbative effects of backreaction, namely the appearance of the BTZ black hole. We now conclude with a discussion of open problems and directions of future research.

\paragraph{Confirming the two-loop correction:} As explained in Section \ref{sec:higher-order}, our formalism provides an efficient method of extracting corrections to the gravitational partition function on thermal $\text{AdS}_3$ to all orders in $G_N$. In particular, we extracted the two-loop correction in \eqref{eq:two-loop-prediction}. As a strong test of our formalism, it would be prudent to verify the two-loop correction through standard perturbation theory. In the gravitational path integral with a massive scalar field, the two-loop correction to the partition function is found by computing two-loop vacuum diagrams. Since pure gravity on thermal $\text{AdS}_3$ is known to be one-loop exact \cite{Maloney:2007ud}, we only need to worry about diagrams which contain graviton-matter interactions. At this order in perturbation theory, there are five such diagrams:
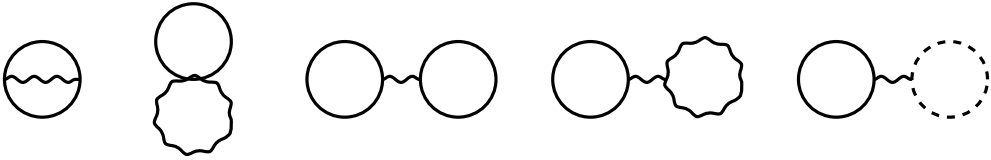
\begin{equation*}
\begin{tikzpicture}[scale = 0.5]
\begin{scope}[xshift = -0.5cm]
\draw[very thick] (0,0) circle (1cm);
\draw[very thick, style={decorate, decoration={snake, amplitude = 0.4mm, segment length = 3mm}}] (-1,0) -- (1,0);
\end{scope}
\begin{scope}[xshift = 3.5cm]
\draw[very thick] (0,1) circle (1cm);
\draw[very thick, style={decorate, decoration={snake, amplitude = 0.25mm, segment length = 3mm}}] (0,-1.1) circle (1cm);
\end{scope}
\begin{scope}[xshift = 9cm]
\draw[very thick] (-1.5,0) circle (1cm);
\draw[very thick] (1.5,0) circle (1cm);
\draw[very thick, style={decorate, decoration={snake, amplitude = 0.4mm, segment length = 3mm}}] (-0.5,0) -- (0.5,0);
\end{scope}
\begin{scope}[xshift = 15.5cm, yshift = 0cm]
\draw[very thick] (-1.5,0) circle (1cm);
\draw[very thick, style={decorate, decoration={snake, amplitude = 0.25mm, segment length = 3mm}}] (1.5,-0.1) circle (1cm);
\draw[very thick, style={decorate, decoration={snake, amplitude = 0.4mm, segment length = 3mm}}] (-0.5,0) -- (0.5,0);
\end{scope}
\begin{scope}[xshift = 22cm, yshift = 0cm]
\draw[very thick] (-1.5,0) circle (1cm);
\draw[very thick, dashed] (1.5,0) circle (1cm);
\draw[very thick, style={decorate, decoration={snake, amplitude = 0.4mm, segment length = 3mm}}] (-0.5,0) -- (0.5,0);
\end{scope}
\end{tikzpicture}
\end{equation*}
where the solid, wavy, and dashed lines correspond to scalar, graviton, and ghost propagators, respectively. The evaluation of these diagrams will be complicated both by the complexity of bulk-to-bulk propagators in AdS as well as the fact that we are working in a thermal background. The techniques developed in upcoming work \cite{BCFLK:2027xxx} should prove useful towards this computation.

\paragraph{Further developing the Virasoro TQFT with matter:} The realization of the $\text{AdS}_3$ gravity Hilbert space in canonical quantization was the primary ingredient in the construction of the `Virasoro TQFT' \cite{Collier:2023fwi}, which has proven an invaluable tool in the computation of gravitational path integrals on complicated three-manifolds. 
By constructing the Hilbert space in a similar manner, in this paper we have provided the first ingredients towards a `TQFT' description of 3D gravity coupled to matter. However there remains much more to develop -- e.g. cutting and gluing rules, and crossing kernels -- in order to fully realize and utilize the system as a TQFT. Of course, such a three-dimensional theory would in no way be a traditional topological field theory (for one, the phase space is infinite-dimensional), but one may still be able to write down formal cutting and gluing rules for computing the gravitational path integral coupled to matter on more complicated three-manifolds. Given the relationship between our analysis and the Wilson spool (see Section \ref{sec:spool}), a lamppost for the construction of a bona fide three-dimensional theory is that it should recover the Wilson spool on arbitrary hyperbolic three-manifolds recently derived in \cite{Bourne:2025azc}.

\paragraph{State counting above the BTZ threshold:}

In Section \ref{sec:black-holes}, we showed that the perturbation series obtained from the path integral on thermal $\text{AdS}_3$ receives nonperturbative corrections from states above the black hole threshold. While the contribution of these nonperturbative effects are divergent in our current formalism, it is still worthwhile to study their implication.

It is interesting to compare the situation in Section \ref{sec:black-holes} to another type of divergence found in 3D gravity. Let $\Sigma_{g,1}$ be a genus $g>0$ Riemann surface with a single asymptotic boundary. The partition function of pure 3D gravity on $\Sigma_{g,1}\times\text{S}^1$ diverges for the same reason as the $n$-particle partition function studied in Section \ref{sec:black-holes} for $n>\lfloor\ell/4G_N\mu\rfloor$. Namely, the fixed-point set of the Hamiltonian has infinite symplectic volume. At the same time, the three-manifolds $\Sigma_{g,1}\times\text{S}^1$ do not admit solutions to Einstein's equations with $\Lambda<0$ if $g>0$. Thus, the divergence of the gravitational partition function may be associated to the fact that the three-dimensional geometry does not admit a hyperbolic metric. It is likely that the same line of reasoning holds in the case with $n$ point particles -- there simply isn't an on-shell metric on thermal $\text{AdS}_3$ which supports $n$ conical defect lines of total angle $n\theta>2\pi$.

In \cite{Maloney:2015ina}, it was argued that, in the case of pure gravity on $\Sigma_{g,1}\times\text{S}^1$, the divergences in the partition function can be avoided if one works with \textit{chiral gravity} instead. This is ultimately because the phase space of chiral gravity on a surface $\Sigma$ is $\mathcal{T}_{\Sigma}/\text{Map}(\Sigma)$, i.e. the moduli space $\mathcal{M}_{\Sigma}$ of curves. Unlike Teichm\"uller space, $\mathcal{M}_{\Sigma}$ admits a compactification that is actually compact (for compact $\Sigma$), resulting in a finite-dimensional Hilbert space on compact surfaces. In \cite{Eberhardt:2022wlc} it was shown explicitly how to obtain finite numbers for the gravitational partition function on topologies of the form $\Sigma_{g,1}\times\text{S}^1$. In the same way, chiral gravity is a natural playground for understanding the nonperturbative computations of Section \ref{sec:black-holes} in a theory where the gravitational Hilbert space is finite-dimensional, even for large particle number. Since the divergences of Section \ref{sec:black-holes} are not present in chiral gravity, it may be possible to recover the black hole entropy \eqref{eq:bh-entropy} through an explicit state-counting computation. This is currently under investigation \cite{Bourne:2027xxx}.

\paragraph{Nonperturbative objects and classical saddles:}

In Section \ref{sec:black-holes} we indicated how the resummation of perturbation theory can tell us about nonperturbative features of scalar backreaction -- namely the free energy of nonperturbative states -- directly from the pole structure of $Z_\text{1-loop}(\bar\mu)$ in the complex $\bar\mu$ plane. As we move down along the real $\bar\mu$ axis the first pole we encounter is at $\bar\mu=0$ which is a simple pole whose residue yields \eqref{eq:BTZpole} and indicates the tantalizing free energy of a BTZ black hole at threshold, i.e. with zero horizon area. An inspection of \eqref{eq:GMY1loops} indicates that there are additional poles to consider beyond $\bar\mu=0$: for every positive integer $\mathsf{m}$, there is a pole at $\bar\mu=-\mathsf{m}$ of order $\mathsf p(\mathsf m)$, the number of partitions of $\mathsf m$. While the residue of each of these poles is complicated by their increasing order, the exponential behavior indicates a free energy of
\beq\label{eq:mysteryF}
    F\sim \frac{1}{8G_N}\left(1+\frac{\mathsf m}{\mu}\right)^2~.
\eeq
As of now, we do not have a solid indication of what objects can contribute free energies of this type. Presumably they correspond to fully backreacted saddlepoints to the Einstein-Hilbert and scalar system. These saddles display a number of interesting features: they have energies lying strictly above the black hole threshold yet possess $O(1)$ entropy, and they appear quantized in inverse powers of the scalar mass and approach a genuine continuum of states for large scalar masses, $\mu\gg1$. One possible hypothesis is that these states correspond to modes of a scalar cloud dressing the threshold BTZ. One piece of guiding intuition is that, by construction, each solution corresponds to a pole in the scalar one-loop partition function on a rigid thermal AdS$_3$ which are also in correspondence with the scalar normal mode frequencies \cite{Denef:2009kn}. One might try to search for a saddle contributing \eqref{eq:mysteryF} by putting a normal mode scalar profile on top of the threshold BTZ and solve the backreaction perturbatively in $G_N$ (in a method similar to \cite{Lifschytz:1993eb,Martinez:1996gn,Casals:2016ioo}), hoping that the backreaction resums. This is morally similar to the constructions of ``quantum black holes'' \cite{Emparan:2020znc,Panella:2024sor} and relatedly one might try to construct them directly in braneworld holography \cite{deHaro:2000wj,Emparan:1999wa,Emparan:1999fd,Emparan:2002px}. We leave an investigation of this to future work.

\paragraph{Generalization to spinning fields:} Another obvious generalization of the formalism developed in this paper is to describe bulk fields with spin. A na\"ive approach to this problem would be to modify the gravitational phase space so that a point particle with helicity $\mathsf{s}$ induces conical defects with different left- and right-moving angles \cite{Li:2024rma}:
\begin{equation}
\theta_L=\frac{8\pi G_N}{\ell}(\mu+\mathsf{s})\,,\quad\theta_R=\frac{8\pi G_N}{\ell}(\mu-\mathsf{s})\,.
\end{equation}
The relationship between defect angles and conformal weights predicts that the corresponding conformal blocks will satisfy
\begin{equation}\label{eq:weight-difference-spin}
h_L-h_R=\mathsf{s}-\frac{4G_N}{\ell}\mu\mathsf{s}\,.
\end{equation}
Invariance of the mapping class group, however, requires that $h_L-h_R\in\mathbb{Z}$, and so the integer quantization of the spin $\mathsf{s}$ seemingly needs to be modified. Assuming that this apparent contradiction can be resolved, one still runs into issues when computing the conformal weights when two particles sit on top of each other. For two particles with helicity $\mathsf{s}_1$ and $\mathsf{s}_2$, the conformal weight of their fusion is
\begin{equation}
h_L(\mathsf{s}_1\otimes\mathsf{s}_2)-h_R(\mathsf{s}_1\otimes\mathsf{s}_2)=\mathsf{s}_1+\mathsf{s}_2-\frac{4G_N(\mu_1+\mu_2)(\mathsf{s}_1+\mathsf{s}_2)}{\ell}\,,
\end{equation}
which is not generically an integer even if \eqref{eq:weight-difference-spin} is.

If one were to naively ignore these issues and formally follow the steps in Section \ref{sec:partition-functions}, one would arrive at the correct one-loop determinant for a spinning field at lowest order in $G_N$ \cite{Giombi:2008vd}. Thus, while there are seemingly obstacles to formulating spinning particles in our framework, we see them as likely temporary confusions as opposed to fundamental obstructions. We leave the resolution of these issues to future work.

\paragraph{Acknowledgements:} We thank Jan de Boer, Alejandra Castro, Geoffrey Compére, Charlie Cummings, Lorenz Eberhardt, Abhijit Gadde, Gaston Giribet, Alexander Maloney, Shiraz Minwalla, Onkar Parrikar, Ronak M Soni, Andy Svesko, Sandip Trivedi, Mykhaylo Usatyuk, and especially Wayne W. Weng for useful discussions. We thank Alejandra Castro, Lorenz Eberhardt, Ronak M Soni, and Wayne W. Weng for comments on a draft of this paper. We additionally acknowledge Alejandra Castro, Albert Law, and Wayne W. Weng for collaboration on related topics. We thank the Galileo Galilei Institute for Theoretical Physics and the organizers of the workshop ``Pathways to Quantum Black Holes: from Effective Theories to Exact Methods'' for their hospitality during the later stages of this work.
JRF thanks TIFR, Kavli IPMU at the University of Tokyo, and New York University for hospitality. RB and BK thank the Université Libre de Bruxelles for hospitality. JRF is supported by FNRS MISU grant 40024018 ``Pushing Horizons in Black Hole Physics.'' RB and BK are supported by STFC consolidated grants ST/T000694/1
and ST/X000664/1. RB was also supported by the Trinity Hall General Graduate Student Fund.

\appendix

\section{Bundles and classes on Teichm\"uller space}\label{app:bundles}

In the main text, we make use of various cohomology classes on Teichm\"uller space. In this appendix, we briefly review their definitions. For a more complete review, see \cite{Zvonkine:xyz}.

\subsection{Cohomology classes on the moduli space}

With or without conical defects, the Teichm\"uller space $\mathcal{T}_{\Sigma}$ of a surface $\Sigma$ is the universal covering space of the moduli space $\mathcal{M}_{\Sigma}$ of complex structures on $\Sigma$. Thus, there exists a projection map $\mathcal{T}_{\Sigma}\to\mathcal{M}_{\Sigma}$. The cohomology classes we consider in the main text are defined by pulling back classes on $\mathcal{M}_{\Sigma}$ under this projection. For simplicity, we assume $\Sigma$ is a compact surface of genus $g$ with $n$ marked points. Away from the boundary points, the moduli space $\mathcal{M}_{g,n,\theta}$ of surfaces with conical defects is isomorphic to the moduli space $\mathcal{M}_{g,n}$ of surfaces with marked points. The latter has a compactification $\overline{\mathcal{M}}_{g,n}$ (the Deligne-Mumford compactification) which differs from the compactification $\overline{\mathcal{M}}_{g,n,\theta}$ that is natural for cone surfaces. The cohomology classes we introduce will be defined for $\overline{\mathcal{M}}_{g,n}$, but can be suitably modified to live on $\overline{\mathcal{M}}_{g,n,\theta}$, see for example \cite{Do:2006xyz,Eberhardt:2023rzz}.

For each $i=1,\ldots,n$, define $\mathbb{L}_i$ to be the line bundle whose fiber over each surface in $\overline{\mathcal{M}}_{g,n}$ is the cotangent space of that surface on the $i^{\text{th}}$ marked point. Sections of $\mathbb{L}_i$ are objects that locally look like $f(z_i)\,\mathrm{d}z_i$ where $z_i$ is the complex coordinate for the marked point. We can define the cohomology classes\footnote{Some care is needed in this definition at the boundary divisors of the moduli space. See \cite{Zvonkine:xyz} for a more complete discussion.}
\begin{equation}
\psi_i=c_1(\mathbb{L}_i)\,.
\end{equation}
The other cohomology class we will need is the Mumford class $\kappa_1$. To define it, we use the forgetful map
\begin{equation}
\pi:\overline{\mathcal{M}}_{g,n+1}\to\overline{\mathcal{M}}_{g,n}
\end{equation}
which forgets the $(n+1)^{\text{th}}$ marked point. Then we define
\begin{equation}
\kappa_m=\pi_{\star}(\psi_{n+1}^{m+1})\,,
\end{equation}
where $\pi_{\star}$ is the pushforward in cohomology (i.e. integration over the fiber of $\pi$). As mentioned above, both the $\psi$ and $\kappa$ classes can be pulled back to cohomology classes on Teichm\"uller space. We will abuse notation and still refer to these pulled-back classes as $\psi_i$ and $\kappa_m$.

\subsection{The pre-quantum line bundle}

The pre-quantum line bundle $\mathcal{L}^k$ used in the quantization of the Teichm\"uller space of cone surfaces has first Chern class
\begin{equation}\label{eq:prequantum-chern-class}
c_1(\mathcal{L}^k)=k\left(\kappa_1-\sum_{i=1}^{n}\alpha_{\theta}^2\psi_i\right)\,.
\end{equation}
Since $\mathcal{T}_{\Sigma,n,\theta}$ is simply connected, this line bundle exists and is unique.

In the main text, we make use of the fact that sections of $\mathcal{L}^k$ transform as conformal blocks of a 2D CFT with central charge $c=24k$ with $n$ operators of conformal weight $h_i=k(1-\alpha_{\theta}^2)$. One way to understand this is as follows. Let $L$ be the line bundle on $\overline{\mathcal{M}}_{g}$ with $c_1(L)=\kappa_1$. It was shown by Friedan and Shenker \cite{Friedan:1986ua} that sections of $L$ transform under Weyl transformations like a CFT conformal blocks with central charge $24$. Thus, sections of $L^{c/24}$ transform with central charge $c$. Let
\begin{equation}
\pi:\overline{\mathcal{M}}_{g,n}\to\overline{\mathcal{M}}_g
\end{equation}
be the map that forgets all mapped points. Since sections of $\mathbb{L}_i^{h_i}$ transform like fields of conformal weight $h_i$, it follows that sections of the line bundle
\begin{equation}
\pi^*L^{c/24}\otimes\mathbb{L}_1^{h_i}\otimes\cdots\otimes\mathbb{L}_n^{h_n}
\end{equation}
transform like conformal blocks for the correlator
\begin{equation}
\Braket{\mathcal{O}_{h_1}(z_1)\cdots\mathcal{O}_{h_n}(z_i)}
\end{equation}
in a CFT of central charge $c$. The first chern class of this bundle is
\begin{equation}\label{eq:conformal-weight-central-charge-class}
\begin{split}
c_1(\pi^*L^{c/24}\otimes\mathbb{L}_1^{h_i}\otimes\cdots\otimes\mathbb{L}_n^{h_n})&=\frac{c}{24}\pi^*\kappa_1+\sum_{i=1}^{n}h_i\psi_i\\
&=\frac{c}{24}\kappa_1+\sum_{i=1}^{n}\left(h_i-\frac{c}{24}\right)\psi_i\,.
\end{split}
\end{equation}
where in the second line we used the pullback formula \cite{Zvonkine:xyz}
\begin{equation}
\pi^*\kappa_1=\kappa_1-\sum_{i=1}^{n}\psi_i\,.
\end{equation}
Comparing to \eqref{eq:prequantum-chern-class}, we find that
\begin{equation}
\mathcal{L}^k=\pi^*L^{c/24}\otimes\mathbb{L}_1^{h_i}\otimes\cdots\otimes\mathbb{L}_n^{h_n}
\end{equation}
with $c=24k$ and $h_i=k(1-\alpha_{\theta}^2)$ (after pulling back to Teichm\"uller space).

\subsection{The canonical bundle}

In Appendix \ref{app:metaplectic_correction}, we will make use of the canonical bundle $\mathscr{K}$ over the moduli space of cone surfaces. We specifically need an expression for the first Chern class of the canonical bundle on the moduli space of cone surfaces. To our knowledge, this has not been computed in the mathematical literature, so we will derive it here. For the purposes of this appendix, we will be more general and consider the moduli space $\overline{\mathcal{M}}_{g,\boldsymbol{\theta}}$ where $\boldsymbol{\theta}=\{\theta_1,\ldots,\theta_n\}$ is a set of potentially different cone angles. The final result is in equation \eqref{eq:c1-canonical-cone}.

To deduce the canonical bundle on the moduli space of cone surfaces, we will use the formalism of \cite{Eberhardt:2023rzz} to relate $\overline{\mathcal{M}}_{g,n,\theta}$ to $\overline{\mathcal{M}}_{g,n}$. As such, we will need the first Chern class of the canonical bundle on $\overline{\mathcal{M}}_{g,n}$. A section of the canonical bundle is an object that is square integrable over the moduli space. From string theory we know that such objects have central charge $26$ and vertex operators with weight $h=1$. This allows us to deduce that, on the interior of the moduli space, we have
\begin{equation}
c_1(\mathscr{K}_{\mathcal{M}_{g,n}})=\frac{1}{12}\left(13\kappa_1-\sum_{i=1}^{n}\psi_i\right)\,.
\end{equation}
The canonical bundle on the compactified moduli space $\overline{\mathcal{M}}_{g,n}$ receives corrections from the boundary class $\Delta$ (the Poincar\'e dual of the compactification divisors) which yields \cite{Harris1982-hf}
\begin{equation}
c_1(\mathscr{K}_{\overline{\mathcal{M}}_{g,n}})=\frac{1}{12}\left(13\kappa_1-\sum_{i=1}^{n}\psi_i-11\Delta\right)\,.
\end{equation}
In the analogy to string theory, the inclusion of the boundary class tells us that string theory partition functions only converge when the worldsheet CFT correlator behaves mildly enough when the worldsheet degenerates.

\begin{figure}
\centering
\begin{tikzpicture}
\draw[very thick] (0,0) to[out = 90, in = 180] (2,2) to[out = 0, in = 180] (5,0) to[out = 180, in = 0] (2,-2) to[out = 180, in = -90] (0,0);
\begin{scope}
\draw[very thick] (2,-2) [partial ellipse = 90-22:90+22:1 and 2];
\draw[very thick] (2,2.8) [partial ellipse = -90-15:-90+15:2 and 3];
\end{scope}
\fill (1,0.5) circle (0.05);
\fill (2,-1) circle (0.05);
\fill (3,0) circle (0.05);
\draw[very thick] (5,0) to[out = 0, in = 180] (7,1) to[out = 0, in = 90] (8,0) to[out = -90, in = 0] (7,-1) to[out = 180, in = 0] (5,0);
\fill (6.5,0.2) circle (0.05);
\fill (7,-0.3) circle (0.05);
\fill (7.5,0) circle (0.05);
\end{tikzpicture}
\caption{A separating divisor in $\overline{\mathcal{M}}_{g,n}$ for which $m$ marked points bubble off into a nodal sphere.}
\label{fig:separating-divisor}
\end{figure}
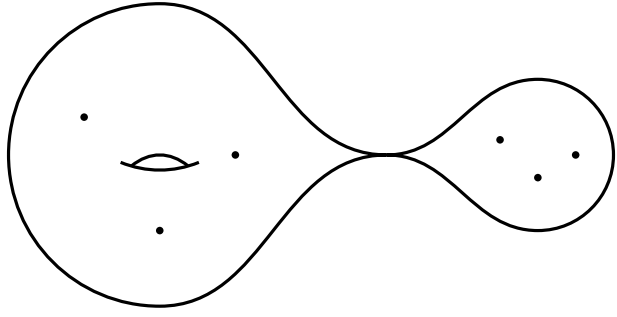

The moduli space $\overline{\mathcal{M}}_{g,\boldsymbol{\theta}}$ is obtained from $\overline{\mathcal{M}}_{g,n}$ as follows. The boundary of $\overline{\mathcal{M}}_{g,n}$ contains components of the form
\begin{equation}
\overline{\mathcal{M}}_{g,n-m+1}\times\overline{\mathcal{M}}_{0,m+1}\subset\overline{\mathcal{M}}_{g,n}
\end{equation}
corresponding to when $m$ marked points approach each other (see Figure \ref{fig:separating-divisor}). Let $I\subset\{1,\ldots,n\}$ be a subset of the marked points and let $\delta_{0,I}$ be the separating divisor for which the points labelled by $I$ bubble off. The moduli space $\overline{\mathcal{M}}_{g,\boldsymbol{\theta}}$ is found by removing the nodal sphere on the right of Figure \ref{fig:separating-divisor} when
\begin{equation}\label{eq:angle-constraint-appendix}
\sum_{i\in I}\theta_i<2\pi\,.
\end{equation}
Furthermore, this step is repeated for every copy of $\overline{\mathcal{M}}_{h,r}$ appearing in any boundary component. This has the effect of `collapsing' the nodal spheres to points when the sum of the defect angles is too small.

Now, since the moduli spaces $\overline{\mathcal{M}}_{g,n}$ and $\overline{\mathcal{M}}_{g,\boldsymbol{\theta}}$ agree everywhere except on the separating divisors $\delta_{0,I}$ with $I$ satisfying \eqref{eq:angle-constraint-appendix}, the canonical bundle on $\overline{\mathcal{M}}_{g,\boldsymbol{\theta}}$ must take the form\footnote{We are using the fairly standard notation that divisors and their Poincar\'e duals are represented by the same characters.}
\begin{equation}
c_1(\mathscr{K}_{\overline{\mathcal{M}}_{g,\boldsymbol{\theta}}})=\frac{1}{12}\Bigg(13\kappa_1-\sum_{i=1}^{n}\psi_i-11\Delta+\sum_{\substack{J\subset\{1,\ldots,n\}\\\sum_{j\in J}\theta_j<2\pi}}r_J\delta_{0,J}\Bigg)\,,
\end{equation}
where $r_I$ are constants to be determined. The separating divisor $\delta_{0,I}$ is a product of $\overline{\mathcal{M}}_{g,n-m+1}$ and $\overline{\mathcal{M}}_{0,m+1}$. Let us define
\begin{equation}
\xi_I:\overline{\mathcal{M}}_{g,n-m+1}\to\overline{\mathcal{M}}_{g,n}
\end{equation}
to be the inclusion map of the left side of Figure \ref{fig:separating-divisor} into the full moduli space. Our consistency condition to determine $r_I$ is that it should reproduce the canonical bundle of the moduli space $\overline{\mathcal{M}}_{g,\boldsymbol{\theta}_I}$ when restricted to the $\overline{\mathcal{M}}_{g,n-m+1}$ component of $\delta_{0,I}$, where
\begin{equation}
\boldsymbol{\theta}_I=\{\theta_i|i\in I^c\}\cup\{\sum_{i\in I}\theta_i\}\,.
\end{equation}
To this end, we use the adjunction formula to compute the canonical bundle on
$\overline{\mathcal{M}}_{g,\boldsymbol{\theta}_I}$. Let $N_I$ be the normal bundle of $\overline{\mathcal{M}}_{g,\boldsymbol{\theta}_I}$ in $\overline{\mathcal{M}}_{g,\boldsymbol{\theta}}$ when $I$ satisfies \eqref{eq:angle-constraint-appendix}. There is a short exact sequence of vector bundles over $\overline{\mathcal{M}}_{g,\boldsymbol{\theta}_I}$ (the conormal exact sequence)
\begin{equation}
0\to N_I^{*}\to \xi_I^*(T^*\overline{\mathcal{M}}_{g,\boldsymbol{\theta}})\to T^*\overline{\mathcal{M}}_{g,\boldsymbol{\theta}_I}\to 0\,,
\end{equation}
the determinant of which yields the isomorphism
\begin{equation}
\mathscr{K}_{\overline{\mathcal{M}}_{g,\boldsymbol{\theta}_I}}\cong \xi_I^*\mathscr{K}_{\overline{\mathcal{M}}_{g,\boldsymbol{\theta}}}\otimes\text{det}N_I~.
\end{equation}
Passing to Chern classes gives
\begin{equation}
c_1(\mathscr{K}_{\overline{\mathcal{M}}_{g,\boldsymbol{\theta}_I}})=\xi_I^*(c_1(\mathscr{K}_{\overline{\mathcal{M}}_{g,\boldsymbol{\theta}}}))+c_1(N_I)\,.
\end{equation}
Thus, the difficulty lies in computing $c_1(N_I)$. Without loss of generality take $I=\{n-m+1,\ldots,n\}$. The normal bundle $N_I$ has rank $m-1$ and is parametrized by infinitesimal separations of the fused conical defect. We can explicitly describe it by keeping one point (say $z_n$) fixed, and allowing the other $m-1$ points to separate from it. The $m-1$ separation modes live in the tangent space to the $i^{\text{th}}$ marked point for $i=n-m+1,\ldots,n-1$, and so we have
\begin{equation}
N_I=\bigoplus_{r=n-m+1}^{n-1} \mathbb{L}^{-1}_r\bigg|_{\overline{\mathcal{M}_{g,\boldsymbol{\theta}_I}}}=\underbrace{\mathbb{L}_{n-m+1}^{-1}\oplus\cdots\oplus\mathbb{L}_{n-m+1}^{-1}}_{m-1\text{ times}}\,.
\end{equation}
Thus, by the splitting principle, the first Chern class of the normal bundle is
\begin{equation}
c_1(N_I)=-(m-1)\psi_{n-m+1}\,,
\end{equation}
and so
\begin{equation}
\begin{split}
c_1(\mathscr{K}_{\overline{\mathcal{M}}_{g,\boldsymbol{\theta}_I}})=\frac{1}{12}\Bigg(&13\kappa_1-\sum_{i=1}^{n-m}\psi_i+(11-r_I-12(m-1))\psi_{n-m+1}\\
&\hspace{3.25cm}-11\Delta+\sum_{J\subset I^c\cup\{n-m+1\}}r_J\delta_{0,J}\Bigg)\,.
\end{split}
\end{equation}
Consistency now requires that $11-r_I-12(m-1)=-1$ for sets with size $|I|=m$. In other words, the consistency of the adjunction formula demands
\begin{equation}
r_I=12(2-|I|)\,,
\end{equation}
so that the Chern class of the canonical bundle is given by
\begin{equation}\label{eq:c1-canonical-cone}
c_1(\mathscr{K}_{\overline{\mathcal{M}}_{g,\boldsymbol{\theta}}})=\frac{1}{12}\Bigg(13\kappa_1-\sum_{i=1}^{n}\psi_i-11\Delta-12\sum_{\substack{J\subset\{1,\ldots,n\}\\\sum_{j\in J}\theta_j<2\pi}}(2-|J|)\delta_{0,J}\Bigg)\,.
\end{equation}
The immediate consequence of this fact is that the pullback of the canonical bundle onto the subset $\overline{\mathcal{M}_{\theta_I}}$ for any $I\subset\{1,\ldots,n\}$ satisfying \eqref{eq:angle-constraint-appendix} is
\begin{equation}
\xi^*_I(c_1(\mathscr{K}_{\overline{\mathcal{M}}_{g,\boldsymbol{\theta}}}))=\frac{1}{12}\left(13\kappa_1-\sum_{i\in I^c}\psi_i+(12|I|-13)\psi_{n-m+1}+\text{boundary data}\right)\,.
\end{equation}
The analogous result for the Teichm\"uller space $\mathcal{T}_{g,\boldsymbol{\theta}}$ is obtained by pulling back under the projection by the mapping class group.

\section{The metaplectic correction}\label{app:metaplectic_correction}

Within the formalism of K\"ahler quantization it is common to make one further adjustment, known as the metaplectic correction. This correction modifies the  original line bundle over phase space through a tensor product with a half-form line bundle as follows. We again consider the quantization of each Teichm\"uller space separately and following \cite{Eberhardt:2022wlc} we denote the canonical bundle over $\mathcal{T}_{\Sigma,n,\theta}$ by $\mathscr{K}$.\footnote{$\mathscr{K}$ stands for ``canonical.''} The contractibility of $\mathcal{T}_{\Sigma,n,\theta}$ ensures the existence of a corresponding half-form bundle $\mathscr{K}^{\frac{1}{2}}$ with the property that
\begin{equation}
    \mathscr{K}^{\frac{1}{2}} \otimes \mathscr{K}^{\frac{1}{2}} = \mathscr{K}~.
\end{equation}
Taking holomorphic sections of $\mathscr{L} = \mathcal{L}^{k} \otimes \mathscr{K}^{\frac{1}{2}}$ we construct the total Hilbert space

\begin{equation}
    \mc H_{\text{grav},n}^0\equiv\text{H}^0\left(\mc T_{\Sigma,n,\theta},\mc L^{k} \otimes \mathscr{K}^{\frac{1}{2}}\right)\otimes\text{H}^0\left(\overline{\mc T}_{\Sigma,n,\theta},\mc L^{-k} \otimes \mathscr{K}^{\frac{1}{2}}\right)~.
\end{equation}
Finally we must check whether this Hilbert space is still consistent with a quotient by the mapping class group $\text{Map}(\Sigma,n)$. In the case of chiral gravity, where we quantize only one holomorphic sector of the $\mc M_\text{grav}$, it is known that a metaplectic correction can not be consistently applied \cite{Eberhardt:2022wlc}. Most clearly this is seen by observing that even in pure gravity a square-root bundle does not exist on $\mathcal{M}_\text{chiral} = \mathcal{T}_{\Sigma}/\text{Map}(\Sigma)$ for $g\geq 3$. In contrast Einstein gravity does admit a consistent metaplectic correction, due to the diagonal action of the mapping class group on $\mathcal{T}_{\Sigma,n,\theta} \times \overline{\mathcal{T}}_{\Sigma,n,\theta}$ \cite{Eberhardt:2022wlc}. 

This correction affects calculations in the quantized theory in three ways. Firstly, the quantized operators acquire an additional term associated to the half-form, although within this work the explicit construction is not required\footnote{In the case of many familiar quantum systems this correction is necessary to find the standard quantized operators, for instance in the harmonic oscillator it is responsible for shifting the ground state energy to $\frac{\hbar\omega}{2}$.}.

Secondly, the inner product of the Hilbert space must be adjusted. $\mathscr{K}^{\frac{1}{2}}$ naturally carries central charge 13 and conformal weight $\frac{1}{2}$ (see Appendix \ref{app:bundles}), and so the holomorphic sections of $\mathscr{L}$ transform as correlation functions \eqref{eq:section_transformation_law} with new parameters
\begin{equation}\label{eq:params_corrected}
    c = 24k + 13~, \qquad\qquad h = k(1-\alpha_{\theta}^2)+\frac{1}{2}~.
\end{equation}
To compensate for this the natural adjustment to the inner product \eqref{eq:npart-Hgm-InnerProd} is to shift the central charge and conformal weights of the timelike Liouville correlator in the opposite fashion. From \eqref{eq:weight_from_angle} we also find that the new relationship between central charge and conformal weight is 
\begin{equation}
    h = \frac{c-1}{24} - k\alpha_\theta^2~,
\end{equation}
which should be compared with the standard expression for the conformal weight in Liouville theory
\begin{equation}
    h = \frac{c-1}{24} + P^2
\end{equation}
in terms of Liouville momentum $P$.

The shift in the central charge is a known one-loop effect in pure gravity \cite{Giombi:2008vd}, while the shift in the conformal weight is less well-known. Writing everything in terms of bulk physical quantities, the shifted conformal weight is simply
\begin{equation}\label{eq:corrected_conformal_weight}
\begin{split}
    h&=\frac{1}{2}+\frac{\mu}{2}-\frac{G_N\mu^2}{\ell}\,.
\end{split}
\end{equation}
Finally, the Chern class of the overall line bundle is corrected. The shift
\begin{equation}
    c_1(\mathscr{L}) = \frac{k}{4\pi}\omega_{\text{WP}} + \frac{1}{2}c_1(\mathscr{K})
\end{equation}
matters both to understand the correction to the conformal weight when multiple defects collide, as well as in performing formal equivariant localization integrals.

\paragraph{Compactification:} We must also consider the effect of the metaplectic correction when $p$ defects collide. In the case where the geometry bubbles into a nodal sphere the extension is straightforward - each defect in the nodal sphere also has corrected conformal weight $h = k(1-\alpha_{\theta}^2)+\frac{1}{2}$. In the case where the defects combine into a single, sharper defect we have that the corrected conformal weight is shifted from \eqref{eq:combined_conformal_weight} by $\frac{p}{2}$ to
\begin{equation}\label{eq:hp-full}
\begin{split}
h(p)&=\frac{p}{2}+\frac{p\mu}{2}-\frac{G_Np^2\mu^2}{\ell}\,.
\end{split}
\end{equation}
This can be seen from the manipulations of Appendix \ref{app:bundles}, specifically by the fact that the line bundle $\mathscr{L}$ restricted to the locus where $p$ points lie on top of each other has the form
\begin{equation}
\begin{split}
\xi_I^*(c_1(\mathscr{L}))=&\frac{24k+13}{24}\kappa_1-\sum_{i=1}^{n-p}\left(k\alpha_{\theta}^2+\frac{1}{24}\right)\\
&-\left(k1-(1-p(1-\alpha_{\theta})^2+\frac{13}{24}-\frac{p}{2}\right)\psi_{n-p+1}+\cdots\,,
\end{split}
\end{equation}
where $\cdots$ represents boundary components. The shifted conformal weight of the fused defect is then read off to be \eqref{eq:hp-full} using the dictionary of equation \eqref{eq:conformal-weight-central-charge-class}.

\paragraph{Implications:} The remainder of the calculation of the thermal AdS$_3$ partition function precedes in much the same way as without the metaplectic correction, leading again to \eqref{eq:thermal-ads3-partition-function}, simply with altered parameters \eqref{eq:params_corrected},\eqref{eq:hp-full}. The shift in $c$ is a gravitational one-loop effect, well known in the literature (e.g. \cite{Giombi:2008vd}). The shift in $h(p)$ further requires a new identification of the worldline coupling parameter to agree with the scalar one-loop result \cite{Giombi:2008vd} as
\begin{equation}
    \mu^2 = m^2 \ell^2 + 1~.
\end{equation}
Such an alteration has a knock-on effect on higher-loop computations, as evidenced in \eqref{eq:two-loop-prediction}. A determination of whether or not the metaplectic correction is consistent with the path integral quantization of a scalar field would therefore most easily be determined by an analogous 2-loop calculation within standard perturbation theory.

The perturbative structure of the theory also appears to be dramatically shifted by the metaplectic correction. Performing again the formal resummation the Gaussian integration kernel is now written most easily in terms of the conformal weight \begin{equation}
    \Delta=1+\mu=1+\sqrt{m^2\ell^2+1}
\end{equation}
as
\begin{equation}\label{eq:corrected_BMF}
    Z(\beta)=\sqrt{\frac{\beta}{8\pi G_N}}\frac{1}{|\Delta-1|}\int_{-\infty}^{\infty}\mathrm{d}\bar\Delta\,e^{-\beta(\Delta-\bar\Delta)^2/(8G_N (\Delta-1)^2)}Z_{\text{1-loop}}(\bar\Delta)\,.
\end{equation}
Note that \eqref{eq:corrected_BMF} has the peculiar feature that the integration kernel limits to the delta function $\delta(\bar\Delta-1)$ when $\Delta\rightarrow 1$. That is to say that the fully backreacted partition function of gravity plus matter with $\Delta=1$ is (perturbatively) {\it one-loop exact}. This is an extremely surprising result that we strongly suspect is not physically sensible. However pressing onwards, we might hope to extract the free energy of non-perturbative states by analyzing the residues of this integral due to poles in $Z_{\text{1-loop}}(\bar\Delta)$. The first such correction no longer corresponds to the standard BTZ threshold $\frac{c}{12}$, but instead occurs at the mass dependent value
\begin{equation}
    \frac{c}{12}(1+\mu^{-1})^2~.
\end{equation}
Should the metaplectic correction be necessary for consistency with the scalar path integral, we would need to make sense of both the one-loop exactness at $\Delta=1$ as well as provide an interpretation for the shifted BTZ threshold.

\bibliography{bibliography}
\bibliographystyle{utphys}

\end{document}